\documentclass[12pt]{article}
\usepackage{amssymb} 
\usepackage{epsfig}
\usepackage{float}
\newcommand{\be}{\begin{equation}}
\newcommand{\ee}{\end{equation}}
\newcommand{\bea}{\begin{eqnarray}}
\newcommand{\eea}{\end{eqnarray}}

\begin{document} 

\begin{center}
{\bf NEUTRINO OSCILLATIONS, MASSES AND MIXING}
\footnote{\em Dedicated to B. Pontecorvo in 90th anniversary of his birth.}
\end{center}
\begin{center}
W. M. Alberico
\end{center}
\begin{center}
{\em Dip. di Fisica Teorica,
Univ. di Torino and INFN, Sez. di Torino, I-10125 Torino, Italy\\}
\end{center}

\begin{center}
S. M. Bilenky 
\end{center}
\vspace{0.1cm} 
\begin{center}
{\em  Joint Institute
for Nuclear Research, Dubna, R-141980, Russia\\}
\end{center}
\begin{center}
{\em INFN, Sez. di Torino and Dip. di Fisica Teorica,
Univ. di Torino, I-10125 Torino, Italy\\}
\end{center}
      
\begin{abstract}
The original B. Pontecorvo idea of neutrino oscillations is discussed.
Neutrino mixing and basics of neutrino oscillations in vacuum and in matter 
are considered.
Recent evidences in favour of neutrino oscillations, obtained in the 
Super-Kamiokande, SNO, KamLAND and other 
neutrino experiments are discussed. Neutrino oscillations 
in the solar and atmospheric ranges of the neutrino mass-squared 
differences are considered in the framework of the minimal scheme with 
the mixing of three massive neutrinos.
Results of the tritium $\beta$-decay experiments and experiments on the
search for neutrinoless double $\beta$-decay 
are briefly discussed.
 
\end{abstract}

\section{Pontecorvo idea of  neutrino oscillations}

B. Pontecorvo came to { the} idea of neutrino oscillations 
in 1957, soon after parity violation in
$\beta$-decay was discovered by Wu et. al.~\cite{WU:57} and the two-component 
theory of massless neutrino was proposed by Landau~\cite{LA:57}, Lee and
Yang~\cite{Lee:57} and Salam~\cite{SA:57}. 

For the first time B.Pontecorvo mentioned the possibility of neutrino 
$\rightleftarrows $ antineutrino transitions in vacuum in his 
paper~\cite{PO:57} on muonium $\rightleftarrows$  antimuonium transition ($\mu^-
e^+ \rightleftarrows \mu^+ e^-$). He believed in { the} analogy 
{ between} the weak 
interaction of leptons and hadrons and looked in the lepton world for
a phenomenon that would be analogous to $K^0 \rightleftarrows
\overline{K}{^0}$ oscillations.
In { his work of 1957}~\cite{PO:57} 
B. Pontecorvo wrote:

\begin{quote}
``If the two-component neutrino theory should turn out to be incorrect
(which at present seems to be rather improbable)
and if the conservation law of neutrino charge would not apply,
then in principle neutrino $\leftrightarrow$ antineutrino transitions
could take place in vacuum.''
\end{quote}

B. Pontecorvo wrote his first paper on neutrino oscillations later in
{ 1958}~\cite{PO2:57} at the time when F. Reines and 
C. Cowan~\cite{RECO:57} { had} just finished the famous reactor experiment, 
which led them to discover the electron { antineutrino}
 through the observation of the inverse $\beta$-process

\begin{equation}
\overline{\nu}_e +p \rightarrow e^+ + n\,.
\label{01}
\end{equation}

At that time R. Davis was doing an experiment with reactor
antineutrinos \cite{DA:57}. R.Davis searched for production of~
$^{37}\rm{Ar}$ in
the process

\begin{equation}
\overline{\nu}_e + \, ^{37}\rm{Cl} \rightarrow e^- + \, ^{37}\rm{Ar}\,,
\label{02}
\end{equation}

\noindent
which is allowed only if the lepton number is not conserved.
A rumor reached B. Pontecorvo that R. Davis had seen some events (2).
B.Pontecorvo, who had earlier been thinking about possible antineutrino
$\rightleftarrows$ neutrino transitions, decided to study this possibility in details.

\begin{quote}
``Recently the question was discussed~\cite{PO:57}  whether there
exist other
{\em mixed} neutral particles beside the $K^0$ mesons, i.e., particles
that differ from the corresponding antiparticles, with the transitions
between particle and antiparticle states not being strictly forbidden.
It was noted that the neutrino might be such a mixed particle, and
consequently there exists the possibility of real neutrino
$\leftrightarrow$ antineutrino transitions in vacuum, provided that
lepton (neutrino) charge is not conserved. 
This means that the
neutrino and antineutrino are {\em mixed} particles, i.e., a symmetric and
antisymmetric combination of two truly neutral Majorana particles $\nu_1$
and $\nu_2$ of different combined parity''
\end{quote}

{ In Ref.}~\cite{PO2:57}  B. Pontecorvo wrote that he has
considered this possibility

\begin{quote}
``\ldots~since it leads to consequences which, in principle, can be
tested experimentally. Thus, for example, a beam of neutral leptons
consisting mainly of antineutrinos when emitted from a nuclear reactor,
will consist at some distance $R$ from the reactor of half neutrinos and
half antineutrinos.''
\end{quote}

In 1958 only { the} electron neutrino was known. In { Ref.}~\cite{PO2:57}
B. Pontecorvo considered oscillations 
of the active right-handed antineutrino to a right-handed neutrino,
the only possible oscillations in the case of one type of neutrino.
In this paper, that was written at { a} time when the two-component
neutrino theory had just appeared and the Davis reactor experiment was not
yet finished, B. Pontecorvo discussed a possible explanation of the Davis
``events''.
Later B. Pontecorvo understood that in the framework of the two-component
neutrino theory, established after M. Goldhaber et al.
experiment~\cite{GGS}, right-handed neutrinos are practically sterile
particles.\footnote{If the lepton number is violated and
neutrinos with definite masses are
Majorana particles the process (\ref{02}) in principle is allowed.
However, the cross section of this process is strongly suppressed by the 
factor $(m/E)^{2}$ ( $m$ is { the} neutrino mass and 
$E$ { the} neutrino energy)}
B. Pontecorvo was in fact the first, who introduced  
the notion of sterile neutrinos in { Ref.}~\cite{PO3}.

  B.Pontecorvo stressed~\cite{PO2:57} that if { the} oscillation length 
$R$ is large, it will be difficult to observe { the} effect of the 
neutrino oscillations in reactor experiments.
He wrote:
\begin{quote}
``Effects of transformation of neutrino into antineutrino and vice versa may be unobservable in the laboratory  because of the large values of R, but will certainly occur, at least, on an astronomical scale.''
\end{quote}

B. Pontecorvo { came} back to the consideration of neutrino oscillations 
in 1967~\cite{PO3}.
At that time { the} phenomenological V-A theory of Feynman and 
Gell-Mann~\cite{FeyGel}, Marshak and Sudarshan~\cite{MarSud} was established 
and in the Brookhaven experiment~\cite{Brookex}, { which} was proposed by 
B.Pontecorvo in 1959~\cite{BP59}, it was proved that (at least) two types on 
neutrinos $\nu_{e}$ and $\nu_{\mu}$ exist in nature. 
B.Pontecorvo confidence in nonzero neutrino 
masses and neutrino oscillations became stronger after these findings.

In { Ref.}~\cite {PO3} B.Pontecorvo
discussed experiments in which 
conservation of lepton numbers were tested
and came to the conclusion that there is ``plenty of room for
violation of lepton numbers, especially in pure leptonic interactions''.
 { He} formulated~\cite{PO2:57,PO3} { the} conditions at which 
neutrino oscillations are possible:
\begin{enumerate}
\item
Lepton numbers, conserved by the usual weak interaction, are violated
by an additional interaction  between neutrinos.
\item
Neutrino masses are different from zero.

\end{enumerate}
In modern terminology these conditions { are equivalent to the assumption} 
that in the total Lagrangian there is a neutrino mass term, non-diagonal 
in flavour neutrino fields.

In { Ref.}~\cite{PO3} B. Pontecorvo considered different lepton numbers 
in the case of two types of neutrinos: additive electron and muon lepton 
numbers $L_{e}$ and $L_{\mu}$, ZKM lepton number~\cite{ZKM} equal to 1 for 
$e^{-}$ and $\mu^{+}$ and -1 for $e^{+}$ and $\mu^{-}$ etc. He stressed
that { if} violation of $L_{e}$ and $L_{\mu}$ { occurs,} in addition 
to the oscillations between active flavour neutrinos 
$\nu_{\mu} \rightleftarrows \nu_{e}$ oscillations between active
and sterile neutrinos   
( $\nu_{\mu} \leftrightarrow \bar \nu_{\mu L}$,
$\nu_{e} \rightleftarrows \bar \nu_{\mu L}$ etc) become possible.

In the case of ZKM lepton number $L$ for one four-component
neutrino $\nu$ we have the following correspondence:
$\nu_{L}\equiv \nu_{e}$, $\nu_{R}\equiv \bar\nu_{\mu}$,
$\bar \nu_{L}\equiv \nu_{\mu}$, $\bar \nu_{R}\equiv \bar \nu_{e}$.
If { the} interaction between neutrinos does not conserve $L$,
 only transitions between active neutrinos
($\nu_{L}\rightleftarrows \bar\nu_{L}$ and 
$\nu_{R}\rightleftarrows \bar\nu_{R}$) are { then} possible.

In { his work of 1967}~\cite{PO3} B. Pontecorvo  discussed 
{ the} oscillations of solar neutrinos. At that time R. Davis started his 
famous experiment on the detection of the solar 
neutrinos. The experiment was based on the radiochemical method, proposed by
B. Pontecorvo in 1946~\cite{BP46}. Solar neutrinos were detected via the 
observation of the Pontecorvo-Davis reaction 

\begin{equation}
\nu_e + \, ^{37}\rm{Cl} \rightarrow e^- + \, ^{37}\rm{Ar}.
\label{03}
\end{equation}

{ Quite remarkably,} in~\cite{PO3} B. Pontecorvo envisaged the solar 
neutrino problem. { Indeed}, before the first results of Davis experiment 
were obtained, he wrote:
\begin{quote}
``From an observational point of view the ideal object is the sun. If
the oscillation length is smaller than the radius of the sun region
effectively producing neutrinos, (let us say one tenth of the sun radius
$R_\odot$ or 0.1 million km for $^8B$ neutrinos, which will give the
main contribution in the experiments being planned now), direct
oscillations will be smeared out and unobservable. The only effect on
the earth's surface would be that the flux of observable sun neutrinos
must be two times smaller than the total (active and sterile) neutrino
flux.''
\end{quote}

It was shown by Gribov and Pontecorvo in 
1969~\cite{GP69} that 
in the neutrino mass term
{ only active left-handed fields $\nu_{e L}$ 
and 
$\nu_{\mu L}$ can enter.}
The corresponding mass term
is called Majorana 
mass term. Particles with definite masses are in this case Majorana 
neutrinos and flavour fields $\nu_{eL}$ and $\nu_{\mu L}$ are connected 
with { the} Majorana fields $\nu_{1L}$ and $\nu_{2L}$ by the mixing relation

\be
\begin{array}{lll}
\nu_e =\cos \theta\,~ \nu_{1L} + \sin \theta\,~ \nu_{2L};\\
&&\\
\nu_{\mu} =-\sin \theta\,~ \nu_{1L} + \cos \theta\,~ \nu_{2L},
\label{04}
\end{array}
\ee
where $\theta $ is the mixing angle.

In { Ref.}~\cite{GP69} { the} expression for { the} survival 
probability of $\nu_e$ was obtained and oscillations of solar neutrinos 
in vacuum were considered.

{ In} the seventies { the} Cabibbo-GIM mixing of $d$ and $s$ 
quarks was established. 
In Ref.~\cite{BP76}
neutrino oscillations in the scheme of the mixing of 
two Dirac neutrinos $\nu_{1}$ and $\nu_{2}$ { were} considered. 
This scheme was based on { the} analogy 
between quarks and leptons. 
{ In the same work} possible values of the mixing angle $\theta $ 
{ were } discussed:

\begin{quote}
``\ldots it seems to us that the special values of mixing angle $\theta
= 0$ (the usual scheme in which muonic charge is strictly conserved) and
$\theta = \pi /4$ are of the greatest interest\footnote{We know from
today's data that 
small and $\pi/4$ values of neutrino mixing angles are really the
preferable ones. Indeed, as we will see later, in the framework of the 
mixing of three massive neutrinos with three mixing angles 
the angle $\theta_{12}$ is small and the angles $\theta_{13}$ and 
$\theta_{23}$ are close to $\pi/4$.}.
"
\end{quote}

Neutrino oscillations and in particular solar neutrino oscillations were 
considered in { Ref.}~\cite{BPnc} in the general case of the Dirac and 
Majorana mass term: { the latter} does not conserve any lepton numbers. 
It is built from active left-handed fields $\nu_{lL}$ and sterile 
right-handed fields $\nu_{lR}$.

In 1978 the first review on neutrino oscillations was published \cite{BP78}.
At that time the list of papers on neutrino oscillations was very short: 
except { for the} cited papers of B. Pontecorvo and his collaborators,
it included { only} references~\cite{MNS,BF,ER,FM,ES} and \cite{BPTb,BP77}.
The review attracted { the} attention of many physicists to the problem of 
neutrino mass and neutrino oscillations.\footnote{It has now about 500 
citations.}  In particular, it was stressed in this review  that due to 
{ the} interference character of neutrino oscillations the investigation 
of this phenomenon is the most sensitive method to search for small neutrino 
mass-squared differences.  

During the { preparation} of the review~\cite{BP78} the authors became 
familiar with the { work of Maki {\em et al.}}~\cite{MNS}, which was 
{ unknown} to many physicists for many years.
In this paper, in the framework of the Nagoya model, in which $p$, $n$ and 
$\Lambda$ were considered as bound states of some vector particle $B^{+}$ and 
leptons, the mixing of two neutrinos { was} introduced.
In addition to { the} usual neutrinos $\nu_{e}$ and $\nu_{\mu}$, which 
they called weak neutrinos, the authors introduced massive neutrinos 
$\nu_{1}$ and $\nu_{2}$, which they called true neutrinos. 
In the Nagoya model { the} proton was considered as a bound state of 
{ the} $B^+$ and { the} neutrino $\nu_1$.

Maki, Nakagawa and Sakata assumed that the fields of weak neutrinos and  of 
true neutrinos are connected by an orthogonal transformation:

\begin{quote}
``It should be stressed at this stage that the definition of the
particle state of the neutrino is quite arbitrary; we can speak of {\em
neutrinos} which are different from weak neutrinos but are expressed by
a linear combination of the latter. We assume that there exists a
representation which defines the true neutrinos through some orthogonal
transformation applied to the representation of weak neutrinos:

\begin{equation}
\begin{array}{lll}
\nu_1 & = & +\nu_e \cos \delta + \nu_\mu \sin \delta \\ 
&&\\
\nu_2 & = & -\nu_e \sin \delta + \nu_\mu \cos \delta \;''
\label{05}
\end{array}
\end{equation}
\end{quote}

In Ref.~\cite{MNS} neutrino oscillations, 
as a phenomenon based on the quantum mechanics of a mixed system, 
were not considered. In connection with the Brookhaven 
experiment~\cite{Brookex}, which was going on at that time, the authors wrote

\begin{quote}
``In the present case, however, weak neutrinos are {\em not stable} due to
occurrence of virtual transmutation $\nu_\mu \rightarrow \nu_e$ induced by
the interaction ${\cal L} = g \overline{\nu}_2
\nu_2 X^+ X$ ($X$ is a field of heavy particles).
If the mass difference between $\nu_1$ and
$\nu_2$, i.e. $|m_{\nu_2} - m_{\nu_1}| = m_{\nu_2} $, is assumed to be a
few
MeV the transmutation time $T\left( \nu_e \leftrightarrow \nu_\mu
\right)$ becomes $\sim$ $10^{-18}$ sec for fast neutrinos with momentum
$\sim$ BeV$/c$.

Therefore a chain of reactions such as

\begin{eqnarray}
\pi^+ & \rightarrow & \mu^+ \nu_\mu  \nonumber \\
\nu_\mu + Z\left(nucleus \right) & \rightarrow & Z^\prime +
      \left( \mu^- \; / \; e^- \right)
\label{07}
\end{eqnarray}
\noindent 
is useful to check the two-neutrino hypothesis 
only when
$|m_{\nu_2} -
m_{\nu_1}| \leq 10^{-6}$ MeV under the conventional geometry of the
experiments. Conversely, the absence
of $e^-$ in the reaction (\ref{07}) will be able
not only to verify two-neutrino hypothesis but also to provide an upper
limit of the mass of the second neutrino $\nu_2$ if the present scheme
should be accepted.''
\end{quote}

Recently very strong, convincing evidence 
in favour of neutrino oscillations, which we will discuss later, were obtained.
It required many years of work and heroic efforts of many
experimental groups to reveal { the} effects of tiny neutrino masses 
and neutrino mixing. From our point of view it is a proper 
time to give a tribute to { the} great intuition of B. Pontecorvo, 
who pursued { the} 
idea of neutrino oscillations for many years at a time when the
general opinion, mainly based on the success of the two-component
neutrino theory, favoured massless neutrinos \footnote{For a collection 
of the papers of B. Pontecorvo see { Ref.}~\cite{BOOK}.}.
The history of neutrino oscillations is an
illustration of the importance of analogy in physics. 
It is also an illustration of { the} importance of new courageous ideas, 
not always in agreement with the general opinion.

\section{Introduction}

Convincing evidence of neutrino oscillations was obtained
in { the} last 4-5 years in experiments with neutrinos from natural sources:
in the atmospheric~\cite {S-K,Soudan, MACRO} and solar neutrino 
experiments~\cite{Cl,GALLEX-GNO,SAGE,S-Ksol,SNO,SNONC,SNOCC}.
Recently strong evidence in favour of neutrino oscillations
was obtained in the reactor KamLAND  experiment~\cite{Kamland}.

The observation of neutrino oscillations { implies}
 that neutrino masses are different from zero and { that the}
states of flavour neutrinos $\nu_{e}, \nu_{\mu}, \nu_{\tau} $
are coherent superpositions (mixtures) 
of the states of neutrinos with definite masses.
All existing data, including astrophysical { ones},
suggest that neutrino masses are much smaller than { the}
masses of leptons and quarks.
The smallness of neutrino masses is a signature of a new, beyond the
Standard Model, physics.

In this review we present the phenomenological theory of neutrino
masses and mixing and the see-saw mechanism of neutrino mass 
generation~\cite{see-saw}. 
We consider neutrino oscillations in vacuum
and transitions between different flavour neutrinos in matter.
Then we will discuss 
the recent experimental evidence in favour of neutrino oscillations:
the results of the Super-Kamiokande 
atmospheric neutrino  experiment~\cite{S-K}, in which a significant 
up-down asymmetry of the high-energy muon events was observed, the results 
of the SNO solar neutrino experiment~\cite{SNO,SNONC,SNOCC},
in which { a} direct evidence for the transition of the solar $\nu_{e}$
into $\nu_{\mu}$ and $\nu_{\tau}$ was obtained and { the} results of
the KamLAND experiment~\cite{Kamland} in which { the} disappearance of the 
reactor $\bar\nu_{e}$ was observed.
We will also discuss the long baseline CHOOZ ~\cite{CHOOZ} 
and Palo Verde~\cite{PaloV} reactor experiments in which no indications
in favour of neutrino oscillations were obtained. Yet
the results of these experiments are very important for neutrino mixing.

Neutrino oscillations in the atmospheric and solar ranges of  neutrino 
mass-squared differences will be considered in the framework of
three-neutrino mixing. We will { show} that 
neutrino oscillations in these two regions are practically decoupled and
are described in the leading approximation by the two-neutrino formulas.

The investigation of neutrino oscillations is based on the following
assumptions, { which are} supported by all existing experimental data.
\begin{enumerate}
\item
{\em The interaction of neutrinos with other particles is described by the 
Standard Model of the electroweak interaction}.

The Standard Charged Current (CC) and Neutral Current (NC)
Lagrangians are given by 
\be
\mathcal{L}_{I}^{\mathrm{CC}}
=
- \frac{g}{2\sqrt{2}} \,
j^{\mathrm{CC}}_{\alpha} \, W^{\alpha}
+
\mathrm{h.c.};
\,~~
\mathcal{L}_{I}^{\mathrm{NC}}
=
- \frac{g}{2\cos\theta_{W}} \,
j^{\mathrm{NC}}_{\alpha} \, Z^{\alpha}
\,.
\label{001}
\ee

Here $g$ is the SU(2)
gauge coupling constant,
$\theta_{W}$ is the weak angle, $W^{\alpha}$ and $Z^{\alpha}$
are { the} fields of { the} charged ($W^{\pm}$) and neutral ($Z^{0}$) 
vector bosons and the leptonic charged current $j^{\mathrm{CC}}_{\alpha}$ and 
neutrino neutral current $j^{\mathrm{NC}}_{\alpha}$ are given by
the expressions:

\begin{equation}
j^{\mathrm{CC}}_{\alpha} =2\, \sum_{l} \bar \nu_{lL} \gamma_{\alpha}l_{L};\,~~
j^{\mathrm{NC}}_{\alpha} =\sum_{l} \bar \nu_{lL}\gamma_{\alpha}\nu_{lL}\,.
\label{002}
\end{equation}
\item

{\em Three flavour neutrinos $\nu_{e}$, $\nu_{\mu}$
and $\nu_{\tau}$ exist in nature}.

From the LEP experiments on the measurement of the width of the  decay
 $Z \to \nu + \bar\nu $ the following value was obtained~\cite{PDG} 
 for the number of flavour neutrinos $n_{\nu_{f}}$:

\begin{equation}
n_{\nu_{f}} = 3.00 \pm 0.06 
\label{003}
\end{equation}

and from the global fit of the LEP data the value

\begin{equation}
n_{\nu_{f}} = 2.984 \pm 0.008.
\label{004}
\end{equation}
was found.
\end{enumerate}

\section{Neutrino masses and mixing}

If neutrino fields enter only in the Standard Model (SM) Lagrangians 
(\ref{001}), neutrinos are massless particles and { the} three flavour 
lepton numbers, electron  ($L_{e}$), muon ($L_{\mu}$)  and tau 
 ($L_{\tau}$),  are separately conserved.
The hypothesis of neutrino mixing is based on the assumption
that in the total Lagrangian there is {\em a neutrino mass term}, which does 
not conserve flavour lepton numbers.
Several mechanisms of the generation of the neutrino mass term were proposed. 
In this section we will discuss the most plausible see-saw 
mechanism~\cite{see-saw}.

{ Let us} start with the phenomenological theory of the neutrino masses and 
mixing.
There are two types of possible neutrino mass terms (see \cite{BPet,BGG}).

\begin{enumerate}
\item

{\em Dirac mass term}
\be
\mathcal{L}^{\mathrm{D}}=- \bar \nu'_{R}\,M^{\mathrm{D}} \nu'_{L}
+\mathrm{h.c.}
\label{005}
\ee
Here

\bea
\nu'_{L}=\left(
\begin{array}{c}
\nu_{e L}\\
\nu_{\mu L}\\
\nu_{\tau L}\\
\vdots
\end{array}
\right);\,~
\nu'_{R}=\left(
\begin{array}{c}
\nu_{e R}\\
\nu_{\mu R}\\
\nu_{\tau R}\\
\vdots
\end{array}
\right)
\label{006}
\eea
and $ M^{\mathrm{D}}$ is a complex non-diagonal matrix.
We assume that in the column $\nu'_{L}$ not only left-handed 
flavour neutrino fields
$\nu_{e L},\nu_{\mu L},\nu_{\tau L} $ but also sterile fields could enter.
Sterile fields do not enter into the standard charged and neutral currents
(see discussion later). 

The matrix $ M^{\mathrm{D}}$ can be diagonalized by { a} bi-unitary 
transformation. We have (see \cite{BPet,BGG}):
\be
M^{\mathrm{D}} = V\,m\,U^{+},
\label{007}
\ee
{ where} $V$ and $U$ are unitary matrices and 
$m_{ik}=m_{i}\,\delta_{ik};\,m_{i}>0$. 
From Eq.~(\ref{005}) and Eq.~(\ref{007}) for the neutrino
 mass term we obtain the standard expression
\be
\mathcal{L}^{D} = - \sum_{i}m_{i}\,\bar \nu_{i}\nu_{i},
\label{008}
\ee 
where $\nu_{i}$ is the field of neutrino with mass $m_{i}$.

The
flavour field $\nu_{lL}$ is connected with  
left-handed field $\nu_{iL}$
by the mixing relation

\be
\nu_{lL} = \sum_{i} U_{li} \nu_{iL}\,.
\label{009}
\ee

It is obvious that the total Lagrangian 
is invariant under { the} global gauge transformation
$$\nu'_{L}\to e^{i\,\alpha}\,\nu'_{L};\, \nu'_{R}\to e^{i\,\alpha}\,\nu'_{R};\,
l\to e^{i\,\alpha}\,l\, $$
where $\alpha$ is an arbitrary constant phase.
From this invariance it follows that the total lepton number 
$ L= \sum_{l}\,L_{l}$ is conserved and $\nu_{i}$ is the field of the Dirac
neutrinos and antineutrinos with $L=1$ for neutrinos and
$L=-1$ for antineutrinos.

\item

{\em Majorana mass term}

\be
\mathcal{L}^{\mathrm{M}}=
-\frac{1}{2}\,( \overline{\nu'_{L}})^{c}\,M^{\mathrm{M}} \nu'_{L}
+\mathrm{h.c.}
\label{010}
\ee

Here 

$$(\nu'_{L})^{c}= C (\bar\nu')^{T}_{L};
\,~~(\overline{\nu'_{L}})^{c}=- (\nu')^{T}_{L}\, 
C^{-1},$$
where $C$ { is  the charge conjugation matrix},
which satisfies the conditions
$$C\,\gamma^{T}_{\alpha}\,~~C^{-1}= -\gamma_{\alpha};\,~C^{T}= -C.$$
The matrix $M^{\mathrm{M}}$ is a {\em symmetrical} matrix. In fact, taking 
into account { the} Fermi-Dirac statistics of the field $\nu'_{L}$, 
we have

$$( \overline{\nu'_{L}})^{c}\,M^{\mathrm{M}} \nu'_{L}=
( \nu'_{L})^{T}\,(C^{-1})^{T}(M^{\mathrm{M}})^{T} \nu'_{L}=
( \overline{\nu'_{L}})^{c}\,(M^{\mathrm{M}})^{T} \nu'_{L}\,.$$

From this relation we obtain

$$M^{\mathrm{M}}=(M^{\mathrm{M}})^{T}\,.$$

For { a} symmetrical matrix we have (see \cite{BPet,BGG})
\be
M^{\mathrm{M}} = (U^{+})^{T}\,m\, U^{+},
\label{011}
\ee
where $U$ is a unitary matrix and 
$m_{ik}=m_{i}\,\delta_{ik};\,~~m_{i}>0$. 

From Eqs.(\ref{010}) and (\ref{011}) it follows that 
the Majorana mass term takes the form
\be
\mathcal{L}^{M} = -\frac{1}{2}\, \sum_{i}m_{i}\,\bar \nu_{i}\nu_{i},
\label{012}
\ee 
where $\nu_{i}$ is the field of neutrino with mass $m_{i}$, which satisfies 
{\em  the Majorana condition}

\be
\nu_{i} =\nu^{c}_{i}= C (\bar\nu)^{T}_{i}.
\label{013}
\ee
The flavour field $\nu_{lL}$
is connected with the Majorana fields $\nu_{iL}$
by the mixing relation
\be
\nu_{lL} = \sum_{i} U_{li} \nu_{iL}.
\label{014}
\ee

In the case of the Majorana mass term
there is no global gauge invariance of the total Lagrangian.
Hence, Majorana neutrinos are truly neutral particles: they do not carry 
{ neither} electric charge { nor} lepton numbers. In other words
Majorana neutrinos and antineutrinos are identical particles.

\end{enumerate}

If there are only flavour fields $\nu_{lL}$ in the column
$\nu'_{L}$, the number of the massive neutrinos
$\nu_{i}$ is equal to the number of flavour neutrinos (three)
and $U$ is a 3$\times$3 Pontecorvo-Maki-Nakagawa-Sakata 
(PMNS)~\cite{PO2:57,PO3,MNS} unitary mixing matrix.

If, in the column $\nu'_{L}$, there are also sterile 
fields $\nu_{s L}$, the number of massive neutrinos $\nu_{i}$ 
will be larger than three. In this case the mixing relation takes the form 

\be
\nu_{lL} = \sum_{i=1}^{3+n_{s}} U_{li} \nu_{iL};\,~
\nu_{s L} = \sum_{i=1}^{3+n_{s}} U_{si} \nu_{iL}\,,
\label{015}
\ee
where $n_{s}$ is the number of the sterile fields, $U$ is a unitary 
$(3+n_{s})\times (3+n_{s})$ matrix, 
$\nu_{i}$ is the field of neutrino with mass $m_{i}$ ($i=1,2,..,3+n_{s}$).

Sterile fields can be right-handed neutrino fields, SUSY fields etc. 
If more than three neutrino masses are small, { the}   
transition of the flavour neutrinos $\nu_{e}$, $\nu_{\mu}$, $\nu_{\tau}$ 
into sterile states becomes possible.

A special interest has the so-called Dirac and Majorana mass term.
Let us assume that, in addition to { the}  flavour fields $\nu_{lL}$, 
in the column $\nu'_{L}$ also { the} fields $(\nu_{lR})^{c}$ 
($l=e,\mu,\tau$) enter.\footnote{ $(\nu_{R})^{c}$ is the left-handed 
component of the conjugated field $\nu^{c}$. In fact, we have 
$$(\nu_{R})^{c}= \frac{1 - \gamma_{5}}{2}\,
C\,\bar\nu{^{T}}=\nu_{L}^{c}. $$}
The Majorana mass term (\ref{010}) can { then} be presented 
in the form of the sum of the left-handed Majorana, Dirac  and
right-handed Majorana mass terms: 

\bea
\mathcal{L}^{\mathrm{D+M}}&=&
-\frac{1}{2}\,( \overline{\nu_{L}})^{c}\,M^{\mathrm{M}}_{L}
 \nu_{L}-  \bar \nu_{R}\,M^{\mathrm{D}}\, \nu_{L}
-\frac{1}{2}\,\bar\nu_{R}\,M^{\mathrm{M}}_{R}
(\nu_{R})^{c} +\mathrm{h.c.} \nonumber\\ 
&=&-\frac{1}{2}\,( \overline{\nu'_{L}})^{c}\,M^{\mathrm{D+M}} \nu'_{L}
+\mathrm{h.c.}
\label{016}
\eea
Here 
\bea
\nu_{L}=\left(
\begin{array}{c}
\nu_{e L}\\
\nu_{\mu L}\\
\nu_{\tau L}
\end{array}
\right),\,~
\nu_{R}=\left(
\begin{array}{c}
\nu_{e R}\\
\nu_{\mu R}\\
\nu_{\tau R}
\end{array}
\right)\,,
\label{017}
\eea
$M^{\mathrm{M}}_{L}$ and $M^{\mathrm{M}}_{R}$
are complex non-diagonal symmetrical 3$\times$3 Majorana matrices and
$M^{\mathrm{D}}$ is a complex non-diagonal 3$\times$3 Dirac matrix.
After the diagonalization of the mass term (\ref{016}) { one gets}

\begin{equation}
\nu_{lL} = \sum_{i=1}^{6} U_{li} \nu_{iL};\,~
(\nu_{lR})^{c} = \sum_{i=1}^{6} U_{\bar l i} \nu_{iL},
\label{018}
\end{equation}
where $U$ is the unitary 6$\times$6 mixing matrix and 
$\nu_{i}$ is the field of the Majorana neutrino with mass $m_{i}$.

The standard see-saw mechanism of the neutrino mass generation~\cite{see-saw}
is based on the assumption that the neutrino mass term
is the Dirac and Majorana one, with $M_{L}^{\rm{M}}=0$.
To illustrate { this} mechanism we will consider the simplest case of one
type of neutrino. Let us assume that the standard Higgs mechanism with one 
Higgs doublet, which is the mechanism of  generation
 of the masses of quarks and leptons, also generates 
the Dirac neutrino mass term
\be
\mathcal{L}^{\mathrm{D}} = -m\,\bar \nu_{R}\nu_{L} +\mathrm{h.c.} 
\label{019}
\ee

It is natural to expect that the mass $m$ is of the same order 
of magnitude 
of the lepton or quark { masses} in the same family. However, we know
from experimental data that neutrino masses are much smaller than the 
masses of leptons and quarks.
In order to ``suppress'' { the} neutrino mass we will assume that there is a
beyond the SM, lepton number violating  mechanism of generation
of the right-handed Majorana mass term

\be
\mathcal{L}^{\mathrm{M}}_{R} = -\frac{1}{2}\,M\,\bar \nu_{R}(\nu_{R})^{c} 
+\mathrm{h.c.}, 
\label{020}
\ee
with $M \gg m$ (usually it is assumed that $M \simeq M_{\rm{GUT}}\simeq
10^{16}\, \rm{GeV}$).
The total mass term is { then} the Dirac and Majorana one, with

\bea
M^{\rm{D+M}}=\left(
\begin{array}{cc}
0&m\\
m&M
\end{array}
\right);\,~
\nu'_{L}=\left(
\begin{array}{c}
\nu_{L}\\
(\nu_{ R})^{c}
\end{array}
\right)
\label{021}
\eea

After the diagonalization of the mass term (\ref{020})
{ one gets}

\bea
\nu_{L} &=& i\,\cos \theta \,
\nu_{1L} +\sin \theta\, \nu_{2L}\nonumber\\ 
(\nu_{R})^{c} &=& -i\,\sin \theta \,\nu_{1L} +\cos \theta\, \nu_{2L}, 
\label{022}
\eea
where $\nu_{1}$ and $\nu_{2}$
are { the} fields of the Majorana particles, with masses

\be
\begin{array}{l}
\displaystyle{
m_{1}= -\frac{1}{2}\,M +\frac{1}{2}\,\sqrt{M^{2}+ 4\,m^{2}}
\simeq \displaystyle{\frac{m^{2}}{M}}\ll m;}\,~\\
\displaystyle{
m_{2}= \frac{1}{2}\,M +\frac{1}{2}\,\sqrt{M^{2}+ 4\,m^{2}}\simeq M.}
\end{array}
\label{023}
\ee
The mixing angle $\theta$ is given by the relation
\be
\tan 2\,\theta = \frac{2\,m}{M}\ll 1\,.
\label{024}
\ee
   
Thus, the see-saw mechanism is based on the assumption
that, in addition to the standard Higgs mechanism of generation of 
the Dirac mass term, there exists a beyond the SM mechanism of 
generation of the right-handed Majorana mass term, which 
changes the lepton number by two and is characterized by a mass
$M\gg m $.\footnote{It is obvious that for charged particles 
such mechanism does not exist.}
The Dirac mass term mixes { the} left-handed field $\nu_{L}$, the component 
of { a} doublet, with a singlet field $(\nu_{R})^{c}$. As a result of this 
mixing the neutrino acquires Majorana mass, which is much smaller than the 
masses of leptons or quarks.

In the case of three generations, in the mass spectrum there { will be} 
three light Majorana masses, much smaller than the masses of quarks and 
leptons, and three very heavy Majorana masses   
of the order of magnitude of the { lepton number violation scale} $M$
(see \cite{AlFer,King,Mohapatra}).

Let us stress that if  neutrino masses are of see-saw origin { then}:
\begin{itemize}
\item
Neutrinos with definite masses are Majorana particles.
\item
There are three light neutrinos.
\item
{ Heavy} Majorana particles must exist.
\end{itemize}

The existence of the heavy Majorana particles, see-saw partners of neutrinos,
could be a source of the barion asymmetry of the Universe 
(see \cite{Buch}).

\section{Neutrino oscillations in vacuum}

In this section we will discuss { the} phenomenon of neutrino oscillations 
(see, for example, \cite{BPet,BGG}). If in the total Lagrangian there is a neutrino mass term the flavour
lepton numbers $L_{e}$, $L_{\mu}$, $L_{\tau}$ are not conserved and transitions between different flavour neutrinos
become possible.

Let us first define 
the states of flavour neutrinos $\nu_{e}$, $\nu_{\mu}$ and $\nu_{\tau}$ in the case of neutrino mixing. 
The flavour neutrinos are particles 
which take part in the standard weak processes. 
For example, the neutrino that is produced together with $\mu^{+}$ in the 
decay $\pi^{+}\to\mu^{+}+\nu_{\mu}$  is the muon neutrino $\nu_{\mu}$;
{ the} electron antineutrino $\bar \nu_{e}$ produces
$e^{+}$ in the process $\bar \nu_{e}+ p \to e^{+}+n $, etc. 

In order to determine the states of the flavour neutrinos let us consider a 
decay
\be
a \to b + l^{+}+ \nu_{l}\,.
\label{025}
\ee
If there is neutrino mixing, the state of the final particles is given by
\be
 |f> = \,~~\sum_{i}|b\rangle \,~ |l^{+}\rangle \,~ 
|\nu_i\rangle \,~ \langle i\,l^{+}\,b\,| S |\,a\rangle \,,
\label{026}
\ee
where $|\nu_i\rangle$ is the state of neutrino with momentum
$\bf p$ and energy 
$$E_i = \sqrt{p^2 + m_i^2 } \simeq p + \frac{ m_i^2 }{ 2 p }; \,~
( p^2 \gg m_i^2 )$$
and $\langle i\,l^{+}\,b\,| S |\,a\rangle $ 
is the element of the S-matrix.

We will assume that the neutrino mass-squared differences are so small 
that { the} emission of neutrinos with different masses can not be 
resolved in the neutrino production (and detection) experiments.
In this case we have

\be
\langle i\,l^{+}\,b\,| S |\,a\rangle \simeq U_{li}^{*}\,~
\langle \nu_{l}\,l^{+}\,b\,| S |\,a\rangle_{SM}\,,
\label{027}
\ee
where 
$ \langle \nu_{l}\,l^{+}\,b\,| S |\,a\rangle_ {SM}$
is the SM matrix element of the process (\ref{025}), calculated under the assumption that all neutrinos $\nu_{i}$ have the same mass.
From (\ref{026}) and (\ref{027}) { we obtain the following expression 
for the normalized state of the flavour neutrino $\nu_l$:}
\begin{equation}
|\nu_l\rangle =
\sum_{i} U_{li}^* \,~ |\nu_i\rangle
\,.
\label{028}
\end{equation}
Thus, in the case of mixing of neutrino fields with
small mass-squared differences, the state of a flavour neutrino
is {\em a coherent superposition (mixture)} of the states of neutrinos 
with definite masses.\footnote{The relation (\ref{028}) is analogous to the 
relation { which connects} the states of $K^{0}$ and $\bar K^{0}$ 
mesons, particles with definite strangeness, with the states of $K_{S}^{0}$ 
and $ K_{L}^{0}$ mesons, particles with definite masses and widths.}

In the general case of active and sterile neutrinos we have

\be
|\nu_{\alpha}\rangle
= \sum_{i} U_{\alpha i}^* \,~ |\nu_i\rangle \,,
\label{029}
\ee
where { the} index $\alpha $ takes the values $e, \mu, \tau, s_{1},...$. 
From the unitarity of the mixing matrix it follows that 
\be
\langle\nu_{\alpha'}|\nu_{\alpha}\rangle = \delta_{{\alpha'}\alpha}.
\label{030}
\ee

The phenomenon of neutrino oscillations is based on the relation (\ref{029}).
Let us consider the evolution of the mixed neutrino states in { the} vacuum.
If at the initial time $t=0$ { a} flavour neutrino $\nu_{\alpha}$
 is produced, the neutrino state at { a} time $t$ { will be}:

\be
|\nu_{\alpha}\rangle_{t}=\,~ e^{-iH_{0}\,t}\,~|\nu_{\alpha}\rangle =
\sum_{i}\,U_{\alpha i}^*\,e^{-iE_it}\,|i\rangle,
\label{031}
\ee
where $H_{0}$ is the free Hamiltonian.

It is important that the phase factors in 
(\ref{031}) for different mass components are different.
Thus, the flavour content of the final state $|\nu_{\alpha}\rangle_{t}$
{ will differ} from the initial one. As we { shall} see later, in spite 
of { the} small neutrino mass-squared differences, the effect of 
{ the} transition of the initial neutrino into another flavour (or sterile) 
neutrino can be large.

Neutrinos are detected through the observation of CC and NC weak processes.
{ By} developing the state $|\nu_{\alpha}\rangle_{t}$ over the total 
system of neutrino states $|\nu_{\alpha}\rangle$, we have

\be
|\nu_{\alpha}\rangle_{t} =\sum_{\alpha'}A(\nu_\alpha \to \nu_{\alpha'})\,
|\nu_{\alpha'}\rangle\,,
\label{032}
\ee
where
\be
A(\nu_\alpha \to \nu_{\alpha'}) = 
 \langle \nu_{\alpha'}\,|e^{-iH_{0}\,t}\,|\nu_{\alpha}\rangle =
 \sum_i U_{\alpha' i}\,~e^{-iE_it}\,~U_{\alpha i}^*
\label{033}
\ee
is the amplitude of the transition $\nu_{\alpha} \to \nu_{\alpha'}$ during 
the time $t$.

Taking into account the unitarity of the mixing matrix, 
{ from (\ref{033}) we obtain the following expression }
for the probability of the transition 
$\nu_{\alpha} \to \nu_{\alpha'}$:\footnote{ We label neutrino masses in 
such a way that $m_1 < m_2 < m_3<...$}

\begin{equation}
{\mathrm P}(\nu_\alpha \to \nu_{\alpha'}) =
\left| \delta_{{\alpha'}\alpha} +\sum_{i\geq 2} U_{\alpha' i}  U_{\alpha i}^*
\,~ (e^{- i \Delta m^2_{i 1} \frac {L} {2E}} -1)\right|^2 \,,
\label{034}
\end{equation}
where $L\simeq t$ is the distance between a neutrino source and a neutrino 
detector, $E$ is { the} neutrino energy and $\Delta m^2_{i 1} = 
m^2_{i}- m^2_{1}$.

Analogously, for the probability of the transition
$\bar\nu_{\alpha} \to \bar \nu_{\alpha'}$ we have:

\begin{equation}
{\mathrm P}(\bar\nu_\alpha \to \bar\nu_{\alpha'}) =
\left| \delta_{{\alpha'}\alpha} +\sum_{i} U_{\alpha' i}^*  U_{\alpha i}
\,~ (e^{- i \Delta m^2_{i 1} \frac {L} {2E}} -1)\right|^2 \,.
\label{035}
\end{equation}

Let us stress the following general properties of the transition
probabilities:

\begin{itemize}

\item Transition probabilities depend on ${L}/{E}$.

\item Neutrino oscillations can be observed if the condition
$$\Delta m^2_{i 1} \frac {L} {E}\gtrsim 1 $$
is satisfied for at least one value of $ i$.

\item From the comparison of { Eqs.}~(\ref{034}) and (\ref{035})
we conclude that the following relation holds
$${\mathrm P}(\nu_\alpha \to \nu_{\alpha'}) =
{\mathrm P}(\bar \nu_{\alpha'}  \to \bar\nu_{\alpha}). $$
This relation is the consequence of the CPT invariance,
intrinsic for any local field theory.

\item
If CP invariance in the lepton sector holds, 
the mixing matrix
$U$ is real in the Dirac case. In the Majorana case the mixing matrix 
satisfies the condition~\cite{BNP}
\be
U_{\alpha i}=U_{\alpha i}^{*}\,~\eta_{i}\,,
\label{036}
\ee
where $\eta_{i}=\pm i$ is the CP parity of the Majorana neutrino $\nu_{i}$.
From { Eqs.}~(\ref{034}), (\ref{035}) and (\ref{036}) we can conclude
that in the case of CP invariance in the lepton sector
we have the following relation
$${\mathrm P}(\nu_\alpha \to \nu_{\alpha'}) =
{\mathrm P}(\bar\nu_\alpha \to \bar\nu_{\alpha'}). $$
\end{itemize}
In conclusion let us introduce { the} parameters which 
characterize { the} 3 $\times$ 3 neutrino mixing matrix. In the general 
$n \times n $ case the unitary matrix is characterized by $n(n-1)/2$ angles 
and  $n(n+1)/2$ phases. 
The phases of lepton fields are arbitrary. If neutrinos with definite masses
are Dirac particles the phases of neutrino fields $\nu_{i}$ are arbitrary
{ as well}. In this case the mixing matrix is characterized by
$$\frac{n(n+1)}{2} -2(n-1)-1 (\rm{common\,~ phase})= \frac{(n-1)(n-2)}{2}$$
physical phases.

Thus, the 3 $\times$ 3 Dirac mixing matrix is
characterised 
by three angles and one phase. Let us parameterize such matrix.
Taking into account the unitary condition $\sum_{i}|U_{ei}|^{2}=1$,
we can choose the first two elements of the first row in following way 
\be
U_{e1} = \sqrt{ 1 -|U_{e 3}|^{2}}\,~ \cos\theta_{12};\,
U_{e2} = \sqrt{ 1 -|U_{e 3}|^{2}}\,~ \sin\theta_{12}\,,
\label{037}
\ee
where $\theta_{12}$ is the mixing angle. The third element of the first 
row can be parameterized as follows

\be
U_{e3} = \sin\theta_{13}
\,~e^{-i\,\delta}, 
\label{037bis}
\ee
where $\delta $ is { the} CP-phase and $\theta_{13}$  the second mixing 
angle. Thus, for $|\nu_e\rangle$ we have

\be
|\nu_e\rangle
=\sqrt{ 1 -|U_{e 3}|^{2}}
\,~ |\nu_{12}\rangle + U_{e3}\,~|\nu_{3}\rangle
\,,
\label{038}
\ee
where 
\be
|\nu_{12}\rangle =\cos\theta_{12}\,~|\nu_{1}\rangle +
\sin\theta_{12}\,~|\nu_{2}\rangle\,.
\label{039}
\ee

The elements of the second and  third line of the mixing matrix 
must be chosen in such a way that the condition

\be
\langle\nu_{l'}|\nu_{l}\rangle = \delta_{l'l}.
\label{040}
\ee
is satisfied.

It is obvious that the two orthogonal vectors
\be
|\nu_{12}^{\perp}\rangle =-\sin\theta_{12}\,~|\nu_{1}\rangle +
\cos\theta_{12}\,~|\nu_{2}\rangle\,,
\label{041}
\ee
\be
|\nu_{e}^{\perp}\rangle =
- U_{e3}^{\star}\,~ |\nu_{12}\rangle +\sqrt{ 1 -|U_{e 3}|^{2}} 
\,~|\nu_{3}\rangle \,,
\label{042}
\ee
are orthogonal to the vector $|\nu_{e}\rangle$.

We have introduced two angles $\theta_{12}$, $\theta_{13}$ and the phase 
$\delta$. The third mixing angle  $\theta_{23}$ will { enter into play} as 
follows 

\be
|\nu_{\mu}\rangle
=
\cos\theta_{23} \,~ |\nu_{12}^{\perp}\rangle +
\sin\theta_{23} \,~|\nu_{e}^{\perp}\rangle
\,,
\label{043}
\ee

\be
|\nu_{\tau}\rangle =
-\sin\theta_{23} \,~ |\nu_{12}^{\perp}\rangle +
\cos\theta_{23} \,~|\nu_{e}^{\perp}\rangle\,.
\label{044}
\ee
From { Eqs.}~(\ref{038}), (\ref{043}) and (\ref{044}) we can obtain all 
the mixing matrix  elements. We shall
be interested in the elements of the  first line and third column. The
former are given by { Eqs.}~(\ref{037}) { and (\ref{037bis}). }
For the elements  $U_{\mu 3}$ and $U_{\tau 3}$ of the third {column} we have

\be
U_{\mu 3} = \sqrt{ 1 -|U_{e 3}|^{2}}\,~ \sin\theta_{23};\,
U_{\tau 3} = \sqrt{ 1 -|U_{e 3}|^{2}}\,~ \cos\theta_{23}\,
\label{045}
\ee

If { the} $\nu_{i}$ are Majorana fields, their 
phases are fixed by the Majorana condition, Eq.(\ref{013}). 
The Majorana mixing matrix can  be presented in the form

\be
U^{M}=U^{D}\,~S(\alpha),
\label{046}
\ee
where $U^{D}$ is the Dirac mixing matrix and 
$S_{ik}(\alpha)= e^{i(\alpha_{i}- \alpha_{1})}\,~\delta_{ik}$ 
is the phase matrix.

The number of  physical phases in the Majorana mixing matrix is equal to
\be
\frac{(n-1)\,(n-2)}{2} +(n-1) = \frac{n(n-1)}{2}
\label{047}
\ee
Thus, { the} 3$\times$3 unitary Majorana mixing matrix is characterized by 
three mixing angles and three CP phases.

From { Eqs.}~(\ref{033}) and (\ref{046})  it is easy to derive the 
following relation { between the amplitudes of the transition
 $\nu_\alpha \to \nu_{\alpha'}$ for the Majorana and Dirac neutrinos:}
\be
A^{M}(\nu_\alpha \to \nu_{\alpha'}) = 
\sum_i U^{M}_{\alpha' i}\,~e^{-iE_it}\,~U_{\alpha i}^{M*}=
\sum_i U^{D}_{\alpha' i}\,~e^{-iE_it}\,~U_{\alpha i}^{D*}=
A^{D}(\nu_\alpha \to \nu_{\alpha'}).
\label{048}
\ee
{ Hence}, the investigation of neutrino oscillations does not allow one to 
distinguish { between} the cases of Dirac or Majorana neutrinos~\cite{BHP}.

\section{Oscillations between two types of neutrinos in vacuum}

We will consider here the simplest case of the transition between two types 
of neutrinos. In this case the index $i$ in Eq.~(\ref{034}) 
takes only one value ($i =2$) and for the transition probability we have 
\be
{\mathrm P}(\nu_\alpha \to \nu_{\alpha'}) =
|\delta_{{\alpha'}\alpha} + U_{\alpha' 2}  U_{\alpha 2}^*
\,~ (e^{- i \Delta m^2 \frac {L} {2E}} -1)|^2 \,,
\label{049}
\ee
where $\Delta m^2= m^2_{2}-m^2_{1} $.

From (\ref{049}) { we obtain the following expression for the 
appearance probability:}
\begin{equation}
{\mathrm P}(\nu_\alpha \to \nu_{\alpha'}) 
= \frac {1} {2} {\mathrm A}_{{\alpha'};\alpha}\,~
 (1 - \cos \Delta m^{2} \frac {L} {2E})\,~~(\alpha' \not= \alpha)\,,
\label{050}
\end{equation}
where 
\begin{equation}
{\mathrm A}_{{\alpha'};\alpha}= 4\,~|U_{\alpha' 2}|^{2}
\,~|U_{\alpha 2}|^{2} = {\mathrm A}_{{\alpha};\alpha'}\,.
\label{050a}
\end{equation}

Let us introduce the mixing angle $\theta$. In the 2$\times$2 case the 
mixing matrix $U$ can be chosen { to be} real. { Hence we set:}
$$U_{\alpha 2} = \sin\theta;\,~U_{\alpha' 2} = \cos\theta. $$

The amplitude ${\mathrm A}_{{\alpha'};\alpha}$ is { then} given by
$${\mathrm A}_{{\alpha'};\alpha} =\sin^{2}  2\theta \,.$$
{ and}  the two-neutrino transition probability takes the standard form
\begin{equation}
{\mathrm P}(\nu_\alpha \to \nu_{\alpha'}) 
=\frac {1}{2}\,~  \sin^{2}  2\theta \,~
 (1 - \cos \Delta m^{2} \frac {L} {2E});\,~~(\alpha' \not= \alpha)
\label{051}
\end{equation}

From Eqs. (\ref{050}) and (\ref{050a}) it is obvious that in the two-neutrino 
case the following relations are valid:
\be
{\mathrm P}(\nu_{\alpha} \to \nu_{\alpha'})=  
{\mathrm P}(\nu_{\alpha'}  \to \nu_{\alpha}) =
{\mathrm P}(\bar\nu_{\alpha} \to \bar\nu_{\alpha'})\,.
\label{052}
\ee
Thus, the CP violation in the lepton sector can not be revealed in the case
of transitions between two types of neutrinos.

The survival probability ${\mathrm P}(\nu_\alpha \to \nu_{\alpha})$
is determined by the condition of conservation of probability:
\be
{\mathrm P}(\nu_\alpha \to \nu_\alpha) 
=1-{\mathrm P}(\nu_\alpha \to \nu_{\alpha'})=
 1 - \frac {1}{2}\,~  \sin^{2}  2\theta \,~
(1 - \cos \Delta m^{2} \frac {L}{2E})\,.
\label{053}
\ee

From { Eqs.}~(\ref{052}) and (\ref{053}) it follows that the 
two-neutrino survival probabilities satisfy the following relation:
\be
{\mathrm P}(\nu_\alpha \to \nu_{\alpha})=  
{\mathrm P}(\nu_{\alpha'}  \to \nu_{\alpha'})\,.
\label{054}
\ee
Thus, { only two oscillation parameters, $\sin^{2} 2\theta$ and  
$\Delta m^{2}$, characterize all transition probabilities 
in the case of a transition} between two types of neutrinos. 

The expressions (\ref{051}) and (\ref{053}) describe periodic transitions 
between two types of neutrinos (neutrino oscillations). They are widely 
used in the analysis of experimental data.\footnote{ As we { shall} see 
later, in the case of three-neutrino mixing neutrino oscillations in 
different ranges of $\Delta m^{2}$ are described, in leading approximation, 
by the two-neutrino formulas.} 

Let us notice that the expression (\ref{051})
 for the two-neutrino transition probability can be { recast} in the form
\be
{\mathrm P}(\nu_\alpha \to \nu_{\alpha'}) 
= \frac {1} {2} \,\sin^{2}  2\theta\,~
 (1 - \cos 2\,\pi \frac {L} {L_{0}})\,,
\label{055}
\ee
where 
\be
L_{0}=4\,\pi \frac {E} {\Delta m^{2} } 
\label{056}
\ee
is the oscillation length.

Finally, the two-neutrino transition probability and the oscillation length
can be rewritten as
\be
{\mathrm P}(\nu_\alpha \to \nu_{\alpha'})
= \frac {1} {2}\,\sin^{2}  2\theta \,~
 (1 - \cos 2.53 \,\Delta m^{2} \frac {L} {E})
\label{057}
\ee
and
\be
L_{0}\simeq 2.48 \,~\frac {E} {\Delta m^{2} }\,~\rm{m} \,,
\label{058}
\ee
where $E$ is the neutrino energy in MeV (GeV), $L$ is the distance in m 
(km) and $\Delta m^{2}$ is { the} neutrino mass-squared difference 
in $\rm{eV}^{2}$.

\section{Neutrino oscillations in matter}

In { the} previous sections we have considered neutrino oscillations 
in { the} vacuum. { However, } if neutrinos pass through the Sun, 
the Earth, supernova, etc.,  matter can significantly { alter the} 
neutrino mixing and the probabilities of the transitions between different 
types of neutrinos. Here we will discuss { the} effects of matter on 
the neutrino transition probabilities (see \cite{W,MS,KP,BGG,Concia}).

The refraction index of a particle with momentum 
$p$ is given by the following classical expression

\be
n  = 1 + \frac {2\pi} {p^2}\,f(0)\, \rho\,,
\label{059}
\ee
{ where} $\rho$ is the number density of matter  and $f(0)$ is the 
amplitude { for } elastic scattering in the forward direction.
The second term in Eq.~(\ref{057})
is due to the coherent scattering of the particle in matter. 

In the case of neutrinos the amplitude of the process $\nu_{e} e\to\nu_{e} e$ 
is different from the amplitude of the processes 
$\nu_{\mu,\tau}e\to \nu_{\mu,\tau}e$. 
This is { related} to the fact that to the matrix element
of the process $\nu_{e} e\to\nu_{e} e$ give contribution diagrams with the
exchange of the $W$ and $Z$ bosons while to the matrix element of the
processes $\nu_{\mu,\tau} e\to \nu_{\mu,\tau} e$ only 
the diagram with the exchange of the $Z$ boson { contributes.}
Thus, { the} refraction indexes of $\nu_{e}$ and $\nu_{\mu,\tau}$ are 
different. { If neutrino mixing occurs,}
this difference { between the} refraction indexes
leads to important effects for the neutrino transitions in matter.

In the flavour representation 
the evolution equation of { a} neutrino with momentum $\bf {p}$ in matter
has the following general form:
\be
i\,\frac{\partial a_{\alpha}(t)}{\partial t} =\sum_{\alpha'}( H_{0}+H_{I})
_{\alpha \alpha'} \,a_{\alpha'}(t).
\label{060}
\ee
Here $H_{0}$ is the free Hamiltonian, $H_{I}$ 
is the effective Hamiltonian of the interaction
of neutrino with matter and $a_{\alpha}(t)$ is the amplitude of the 
probability to find $\nu_{\alpha}$ at { a} time $t$. We have
\be
(H_{0})_{\alpha \alpha'} =<\alpha|\,\rm{H_{0}}\,|\alpha'> =
\sum_{i}U_{\alpha i}\,E_{i}\, U^{*}_{\alpha'i}\,,
\label{061}
\ee
where $U$ is the neutrino mixing matrix in vacuum and  
$E_{i}\simeq p+ \frac{m_{i}^{2}}{2\,p}$.

Let us consider the effective Hamiltonian of the interaction of 
neutrinos with matter. We will limit ourselves to the case 
of the flavour neutrinos.

{ Due to the} $\nu_{e}-\nu_{\mu}-\nu_{\tau}$ universality of the NC, 
the NC part of the effective Hamiltonian is proportional to the unit matrix. 
This part of the Hamiltonian can be excluded from the evolution equation 
{ through a} phase transformation of the function $a(t)$.
Thus, we need to take into account only the CC part of the Hamiltonian of 
$\nu_{e}-e$ interaction. We have:
\be
(H_{I})_{e e} = <\nu_{e}\, \rm{mat}|\,\frac{G_F}{\sqrt{2}}\, 
2\, \overline{\nu}_{eL} \gamma^\alpha \nu_{eL}\,~ 
\overline{e} \gamma_\alpha (1 - \gamma_5) e\,|\nu_{e}\, \rm{mat}> 
\simeq \sqrt{2} G_F\, \rho_e \,,
\label{062}
\ee
where $|\nu_{e}\, \rm{mat}>$ is the state vector of { a} neutrino with 
momentum $\bf{p}$ and matter and
$\rho_e$ is the electron number density.\footnote{The same expression for 
the effective Hamiltonian of neutrinos in matter can be obtained from 
{ Eq.}~(\ref{059}) (see \cite{W})}.

We will consider now the simplest case of two flavour neutrinos.
The  mixing matrix has the form 

\begin{equation}
U = \left(\begin{array}{rr} \displaystyle
\cos\theta\null & \null \displaystyle \sin\theta
\\ \displaystyle - \sin\theta \null 
& \null \displaystyle \cos\theta
\end{array}\right)
\label{063}
\end{equation}
where $\theta$ is the {\em vacuum} mixing angle.  
From Eqs.~(\ref{061})-(\ref{063}) the total effective Hamiltonian 
of neutrinos with momentum p in matter { can then be written 
as:}\footnote{We omit { the irrelevant} unit matrix.}
\begin{equation}
H = \frac {1} {4p}\left(
\begin{array}{rr} \displaystyle
- \Delta m^2 \cos 2 \theta + A
\null & \null \displaystyle
\Delta m^2 \sin 2 \theta
\\ \displaystyle
\Delta m^2 \sin 2 \theta
\null & \null \displaystyle
\Delta m^2 \cos 2 \theta - A
\end{array} \right)
\label{064}
\end{equation}
{ where }
$$\Delta m^2 = m^2_2 - m^2_1;\,~~ A = 2 \sqrt{2} G_F \rho_e\, p.$$

The total Hamiltonian (\ref{064}) can be easily diagonalized. We have 
\be
H = U^m\, E^m\, U^{m +}\,,
\label{065}
\ee
where $E^m_i$ ($i=1,2$) are { the} eigenvalues of the matrix $H$ and
\be
U^m =  \left(
\begin{array}{rr} \displaystyle
\cos \theta^m \null & \null \displaystyle
\sin\theta^m \\ \displaystyle
- \sin\theta^m \null & \null \displaystyle
\cos \theta^m \end{array} 
\right)
\label{066}
\ee
is the orthogonal { mixing matrix in matter}.

For the eigenvalues $E^m_{1,2}$ we find the following expressions:
\be
E^m_{1,2} = \mp \frac {1} {4p} \sqrt{(\Delta m^2\cos\, 2\theta - A)^2 + 
(\Delta m^2 \sin\,  2\theta)^2)},
\label{067}
\ee
{ while} the mixing angle $\theta^m$ is given by the relations
\bea
&&\tan\,  2\theta^m = \frac {\Delta m^2 \sin\,  2\theta} 
{\Delta m^2 \cos\,  2\theta - A};\,~~
\nonumber\\
&&\cos\,  2\theta^m = \frac {\Delta m^2 \cos\, 2\theta - A} 
{\sqrt{(\Delta m^2 \cos\,  2\theta - A)^2 + (\Delta m^2 \sin\,  2\theta)^2}}\,.
\label{068}
\eea

Let us consider { now} the simplest case of  constant electron density.
The probability of the transition $\nu_{\alpha}\to\nu_{\alpha'}$ 
($\alpha\not=\alpha'$) in  matter  is  given in this case by the expression
\be
P^m(\nu_{\alpha} \to \nu_{\alpha'}) =P^m(\nu_{\alpha'} \to \nu_{\alpha})=
\frac {1} {2} \sin^2 2\theta ^m\, (1-\cos\Delta E^m\, L)\,,
\label{069}
\ee
where $L $ is the distance which { the} neutrino { travels} in matter and
\be
\Delta E^m = E_2^m - E_1^m = \frac {1} {2p}\,
\sqrt{(\Delta m^2\,  \cos\, 2 \theta - A)^2 + 
(\Delta m^2 \sin\,  2 \theta )^2}\,.
\label{070}
\ee
From { Eqs.}~(\ref{069}) and (\ref{070}) it follows that the oscillation 
length in matter is given by
\be
L_{0}^{m}=\frac{4\pi \,p}{\sqrt{(\Delta m^2\,  \cos 2 \theta - A)^2 + 
(\Delta m^2 \sin\,  2 \theta )^2}}\,.
\label{071}
\ee
By putting $\rho_{e}=0$, we can easily see that the expressions (\ref{069}) and
(\ref{071}) { reduce to the} corresponding vacuum expressions 
(\ref{055}) and (\ref{056}). Again the survival 
probability of $\nu_{\alpha}$ ($\nu_{\alpha'}$)
is given by the condition of  conservation of the { total} probability:  

\be
P^m(\nu_{\alpha} \to \nu_{\alpha}) =
P^m(\nu_{\alpha'} \to \nu_{\alpha'}) =
1 - P^m (\nu_{\alpha} \to \nu_{\alpha'}).
\label{072}
\ee

The neutrino mixing angle and oscillation length in matter 
can be significantly different from the vacuum values.
Let us assume that the condition 

\be
\Delta m^2 \cos\, 2\theta = A = 2\, \sqrt{2}\, G_F\, \rho_e\, p
\label{073}
\ee
is satisfied. { Then  from (\ref{068}) it follows that} the mixing in 
matter is maximal ($\theta^m = \pi/4$), independently on the value of the 
vacuum { mixing} angle $\theta$. In this case the oscillation length in 
matter { turns out to be}:

$$L_{0}^{m}= \frac{L_{0}}{\sin\, 2\theta}\,$$
where $L_{0}$ is the vacuum oscillation length.

In  general  the electron density is not a constant. For example, 
in the Sun the density is maximal in the center of the Sun and 
it decreases practically exponentially to its periphery.

{ Hence} in the general case the evolution equation of 
neutrinos in matter has the form
\be
i \frac {\partial a(t)} {\partial t} = H^{m}(t)\, a(t)\,,
\label{074}
\ee
where $H^{m}(t) =H_{0}+ H_{I}(t)$ is the total effective Hamiltonian 
in the flavour representation.

This equation can be easily solved in the case of slowly changing 
density of electrons $\rho_{e}$. In fact, the hermitian  Hamiltonian 
$H^{m}(t)$ can be diagonalized by the unitary transformation

\be
H^{m}(t) = U^m(t)\, E^m(t)\, U^{m+}(t),
\label{075}
\ee
where $U^m(t) U^{m+}(t) =1$ and $E_i^m(t)$  is the
eigenvalue of the Hamiltonian. Let us now introduce the function
\be
a'(t)=U^{m+}(t)\,a(t);
\label{076}
\ee
from { Eqs.}~(\ref{074})-(\ref{076}) we have
\be
i \, \frac {\partial a'(t)} {\partial t} = 
(E^m(t) - i\, U^{m+}(t)\,  \frac {\partial U^m(t)} {\partial t})\, a'(t)\,.
\label{077}
\ee
{If we} assume that the function  $\rho_e(t)$ depends so weakly 
on $t$ that we can neglect the second term in Eq.~(\ref{077}), { then }
 in this approximation ({ the so-called} adiabatic approximation) 
the solution of the evolution equation is { obviously} given by
\be
a'(t) = e^{- i \int_{t_0}^t{E^m(t) dt}}\, a'(t_0),
\label{078}
\ee
where $t_0$ is { the} initial time.

From { Eqs.}~(\ref{076}) and (\ref{078}) for the solution 
of the evolution equation in the flavour representation we have
\be
a(t) = U^m(t)\, e^{- i \int_{t_0}^t{E^m(t) dt}}\, U^{m+}(t_0)\,a(t_0)\,.
\label{079}
\ee
Hence, in the adiabatic approximation, the amplitude of the transition 
$\nu_\alpha \to \nu_{\alpha'}$ during { a time interval} ($t-t_{0}$) 
is given by the expression
\be
A(\nu_{\alpha}\to \nu_{\alpha'}) = \sum_{i} U^m_{{\alpha'}i}(t)\,
e^{- i \int_{t_0}^t{E_i^m(t) dt}}\, U^{m*}_{{\alpha}i}(t_0)\,.
\label{080}
\ee

The latter is similar in form to the expression (\ref{033})
for the amplitude of transitions in vacuum: this is connected with the fact
that in the adiabatic approximation neutrino remains on the same energy
level. The expression (\ref{080}) has { a} simple meaning:
$U^{m*}_{{\alpha}i}(t_0)$
is the amplitude of the transition from the state of the initial $\nu_{\alpha}$
to the state with energy $E_{i}(t_0)$ ;
the factor $e^{- i \int_{t_0}^t{E_i^m(t) dt}}$ describes
the propagation in the state with definite energy; 
$U^m_{{\alpha'}i}(t)$ is the amplitude of the transition from the state
with energy $E^{m}_{i}(t)$ to the state of the final 
$\nu_{\alpha'}$. The coherent sum over all $i$ must be performed.

The transition probability must be averaged over the 
the region where neutrinos are produced, over { the} energy resolution, etc.
Oscillatory terms in the transition probability usually disappear
after the averaging. { In this case from Eq.~(\ref{080}) 
we obtain}  ($x\simeq t$):
\be
P(\nu_\alpha \to \nu_{\alpha'}) = \sum_{i} 
|U^m_{{\alpha'}i}(x)|^2\, |U^m_{{\alpha}i}(x_0)|^2
\label{081}
\ee
{ Hence}, in the adiabatic approximation the averaged transition
probability is determined by the elements of the mixing 
matrix in matter at the initial and final points.

{ In} the case of two neutrino flavours, from { Eqs.~}(\ref{066}) 
and (\ref{081}) we obtain the following expression for the $\nu_e$ 
survival probability:
\be
P(\nu_e \to \nu_e) 
= \frac {1} {2} \left[1 +\ cos\, 2\theta^m(x)\, \cos\, 2\theta^m(x_0)\right]\,.
\label{082}
\ee

Let us now consider  solar neutrinos 
in the two-neutrino case. If at some point $x =x_{R}$  
the MSW \cite{W,MS} resonance condition
\be
\Delta m^2\, \cos\, 2\theta = A (x_{R})
\label{083}
\ee
is satisfied, { neutrino} mixing at this point is maximal
($ \theta(x_{R}) = \pi/4$). 

If the resonance condition is fulfilled,  $cos 2\theta > 0$ 
(neutrino masses are labeled in such a way that $\Delta m^2 > 0$ ).
Let us assume that the resonance point is in the region between the 
Sun core, where neutrinos are produced, and the surface of the Sun:
then at the initial point $x_0$ the electron density will be larger than at the 
resonance point $x_{R}$ and we have $A (x_{0})> \Delta m^2 cos 2\theta$.  
In this case from Eq.~(\ref{068}) it follows that 
$\cos\, 2\theta^m(x_{0})<0$ and Eq.~(\ref{082}) predicts:
$$P(\nu_e \to \nu_e) < \frac{1}{2}\,.$$

If the condition

\be
A(x_0) >> \Delta m^2
\label{084}
\ee
is satisfied, { then} from { Eq.}~(\ref{068}) we have
\be
\cos\, 2\theta^m(x_0) \simeq - 1. 
\label{085}
\ee
Taking into account that $\theta^m(x)=\theta$ on the surface of the sun, 
for the $\nu_{e}$ survival probability we obtain the following relation
\be
P(\nu_e \to \nu_e) \simeq \frac {1} {2} (1 -  \cos\, 2\theta)\,.
\label{086}
\ee

This relation implies that if the vacuum angle $\theta$ is small, the $\nu_e$
survival probability is close to zero { and}  practically
all $\nu_e$'s are transfered into other neutrino states.  

The MSW resonance condition (\ref{083}) was written in units 
$\hbar = c = 1$. It can be rewritten as follows
\be
\Delta m^2\, \cos\, 2\theta \simeq 0.7 \cdot 10^{-7}\,E\, \rho\,~ \rm{eV}^2,
\label{087}
\ee
 where $\rho$ is the density of matter in g $\cdot$ cm$^{-3}$ and
$E$ is neutrino energy in MeV. 
In the central region of the Sun $\rho \simeq 10^2\,{\rm g}\cdot{\rm cm}^{-3}$; 
the energy of  solar neutrinos is $E \simeq 1\,\rm{ MeV}$. Thus, { the} 
resonance condition is satisfied for the solar neutrinos if 
$\Delta m^2 \simeq 10^{-5}\rm{eV}^2$.

In the general case of non-adiabatic transitions, for the averaged 
transition probability we have
\be
P(\nu_\alpha \to \nu_{\alpha'}) = 
\sum_{i,k} |U^m_{{\alpha'}k}(x)|^2\, P_{ki}\, |U^m_{{\alpha}i}(x_0)|^2
\label{088}
\ee
where $P_{ki}$ is the probability of transition from the energy level $i$ 
to the energy level $k$.

{ If we limit ourselves to the} two-neutrino case, 
from the conservation of the total probability we have
\be
P_{11} = 1- P_{21};\,
P_{22} = 1 - P_{12};
\label{089}
\ee
taking into account that $P_{21} = P_{12}$,
from { Eqs.}~(\ref{066}), (\ref{088}) and (\ref{089}) we obtain the 
following general expression for the $\nu_e$ survival probability~\cite{Parke}
\be
P(\nu_e \to \nu_e) =
\frac {1} {2} + \left(\frac {1} {2} - P_{12}\right)\, 
\cos\, 2\theta^m(x)\, \cos\, 2\theta^m(x_0)\,.
\label{090}
\ee

Different approximate expressions for 
the transition probability $P_{12}$ exist in { the} literature.
 In the Landau-Zenner approximation, which is based on the assumption that transitions 
occur mainly in the resonance region, we have
\be
P_{12} = e^{- \displaystyle{\frac {\pi} {2}\, \gamma_R\, F}}\,,
\label{091}
\ee
where
\be
\gamma_R = 
= \frac{\Delta m^2 \sin^2\, 2\theta} {2 p\,\ cos\, 2\theta\, 
\left|\displaystyle{\frac{d}{dx}}\,\ln\, \rho_e(x_R)\right| }\,.
\label{092}
\ee
For { an} exponential density $F = 1 -\tan^2\theta$; for { a} 
linear density $F = 1$. Let us notice that the adiabatic approximation 
is valid if $\gamma_R >> 1$. In this case $P_{12} \simeq 0$.

\section{Neutrino oscillation data}

\subsection{Evidence in favour of oscillations of atmospheric neutrinos}

Atmospheric neutrinos are mainly originated from the decays of charged pions 
and consequent decays of muons:

\be
\pi \to \mu  + \nu_{\mu};\,~~
\mu  \to e + \nu_{\mu} + \nu_{e}\,, 
\label{093}
\ee
Pions are produced in the processes of interaction of the cosmic rays in 
the atmosphere. In the Super-Kamiokande (S-K) experiment~\cite{S-K} 
electron and muon neutrinos were detected via the observation of the 
Cherenkov light 
in the large water-Cherenkov detector (50 kt of $\rm{H_{2}\,O}$).

At relatively small energies ($\lesssim 1\rm{ GeV}$ ) practically
all muons decay in the atmosphere and from (\ref{093}) it follows that 
the ratio of the numbers of muon and electron events $R_{\mu/e}$
must be equal to 2 (if there are no neutrino oscillations).
At higher energies the expected ratio $R_{\mu/e}$ is 
larger than two. It can be predicted, however, with an accuracy better 
than 5 \%.

The ratio $R_{\mu/e}$, measured in the S-K~\cite{S-K} and 
SOUDAN2~\cite{Soudan} atmospheric neutrino experiments 
is significantly smaller than the ratio $(R_{\mu/e})_{\rm{MC}}$, predicted  
under the assumption of no neutrino oscillations. In the S-K experiment for 
the ratio of ratios in the sub-GeV region ($E_{\rm{vis}}\leq 1.33\,\rm{GeV}$) 
and multi-GeV  region ($E_{\rm{vis}}> 1.33\,\rm{GeV}$) the values
$$\frac{(R_{\mu/e})_{meas}}{(R_{\mu/e})_{\rm{MC}}} 
= 0.638 \pm 0.016 \pm 0.050;\,~\frac{(R_{\mu/e})_{meas}}{(R_{\mu/e})_{\rm{MC}}}
= 0.658 \pm 0.030 \pm 0.078$$
were obtained, respectively.

The fact that the ratio $(R_{\mu/e})_{meas}$ is significantly
smaller than the predicted ratio $(R_{\mu/e})_{\rm{MC}}$
was known from the results of the previous atmospheric neutrino experiments, 
Kamiokande~\cite{Kam} and IMB~\cite{IMB}. For many years
this ``atmospheric neutrino anomaly''  was considered as an indication in 
favour of { the} disappearance of muon neutrinos due to neutrino 
oscillations.

\begin{figure}[t]
\begin{center}
\includegraphics*[bb=45.1211 300.4404 549.3418 799.3911, width=0.7\textwidth]{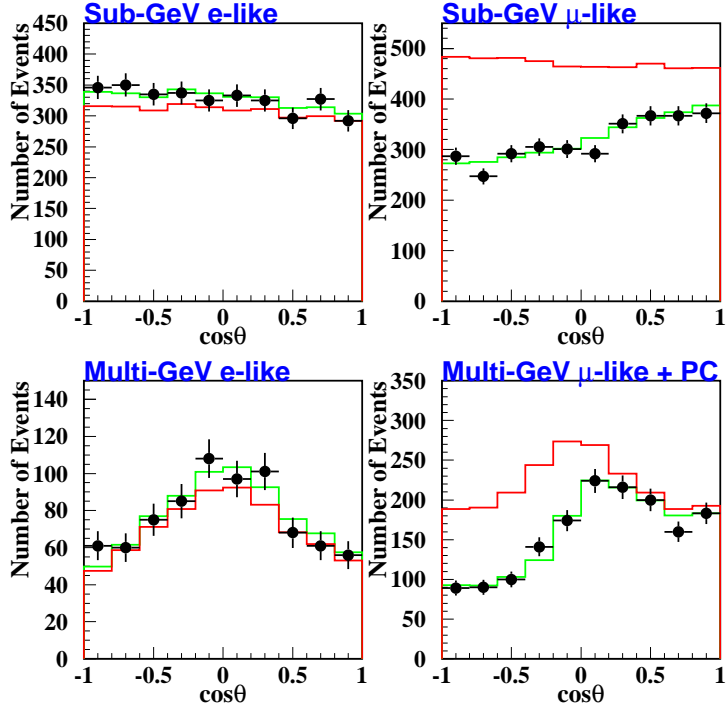}
\end{center}
\caption{ \label{sk-nakaya-0209036-f1}
Zenith angle distribution of Super-Kamiokande sub-GeV and multi-GeV
electron  and muon events.
The black histogram shows the Monte Carlo prediction
(under the assumption of no neutrino oscillations).
The gray histogram was obtained for $\nu_\mu\to\nu_\tau$
oscillations with the best-fit values 
$\Delta m^2 = 2.5 \times 10^{-3} \, \mathrm{eV}^2$
and $\sin^22\vartheta=1.0$.
Figure taken from Ref.~\cite{Nakaya:2002ki}.
}
\end{figure}

The compelling evidence in favour of neutrino oscillations has been 
obtained recently by the S-K collaboration~\cite{S-K}, from the observation 
of the large up-down asymmetry of the atmospheric high energy muon events.
In the S-K experiment \cite{S-K} the zenith angle dependence of the numbers of 
electron and muon events was measured. { In the absence of} neutrino 
oscillations,  the number of the electron (muon) events in the multi-GeV 
region { must obey} the following relation:\footnote{For the multi-GeV 
events the effect of the magnetic field of the Earth can be neglected.}
\be
N_{l}(\cos\theta_{z})= N_{l}( -\cos \theta_{z})\,~~ (l=e,\mu)\,,
\label{094}
\ee
where $\theta_{z}$ is the zenith angle.

The results of the S-K experiment on the measurement of the zenith-angle 
distribution of the atmospheric neutrino events is { shown} in 
Fig.~\ref{sk-nakaya-0209036-f1}. As it is seen from this figure, for 
the electron events there is a good agreement between the data and 
relation (\ref{094}). Instead, for the Multi-GeV muon events a 
significant violation of relation (\ref{094}) was observed.
The ratio of the number of up-going muons 
($-1 \leq \cos\theta_{z}\leq -0.2$) to the number of down-going muons 
($0.2 \leq \cos\theta_{z} \leq 1$) was found to be
$$  \left(\frac{U}{D}\right)_{\mu}= 0.54 \pm 0.04 \pm 0.01.$$
At high energies the direction of leptons practically coincides with the   
direction of neutrinos.
The up-going muons are produced by neutrinos which travel distances
from $\simeq 500\,\rm{km}$ to  $\simeq 13000 \, \rm{km}$ and  
the down-going muons are produced by neutrinos which travel distances 
from $\simeq 20\, \rm{km}$ to $\simeq 500\, \rm{km}$.
{ Hence} the observation of the up-down asymmetry 
clearly demonstrates the dependence of the number of  muon neutrinos
on the distance which they travel from the production point in the 
atmosphere to the detector.

The S-K data \cite{S-K} and { the} data of { the} other atmospheric 
neutrino experiments, SOUDAN 2~\cite{Soudan} and MACRO~\cite{MACRO}, 
are perfectly described if we assume that the two-neutrino $\nu_{\mu}\to
\nu_{\tau}$ oscillations take place. From the analysis of the S-K data 
it was found~\cite{S-K} that at 90 \% $\rm{CL}$ { the}  
neutrino oscillation parameters 
$\Delta m^{2}_{atm}$ and $\sin^{2}2 \theta_{atm}$ 
are in the range
$$1.6\cdot 10^{-3}\leq  \Delta m^{2}_{atm}\leq 3.9 \cdot 10^{-3}\,~\rm{ eV}^{2};\,~ \sin^{2}2 \theta_{atm}>0.92;\, ,$$
the best-fit values of the parameters { being} equal to
\be
\Delta m^{2}_{atm}=2.5\cdot 10^{-3}\rm{ eV}^{2};\,~\sin^{2}2 \theta_{atm}=1.0 
\,~(\chi^{2}_{\rm{min}}= 163.2/ 170\,\rm{d.o.f.}) 
\label{095}
\ee

\subsection{ Indications in favour of neutrino oscillations obtained in the
K2K experiment}

Neutrino oscillations in the atmospheric range of $\Delta m^{2}$
are investigated in the first long
baseline accelerator experiment K2K~\cite{K2K}. 
In this experiment neutrinos, originated mainly  from the decay of
pions, produced at { the} 12 GeV KEK accelerator,
are recorded by the S-K detector at { a} distance of about 250 km from the
accelerator. The average neutrino energy is $\simeq$ 1.3 GeV.

Two near detectors at { a} distance of about 300 m 
from the beam-dump target are used in the K2K experiment: { a} 1 kt 
water-Cherenkov detector and { a} fine-grained detector. 
The total number  and spectrum of muon neutrinos, observed in the S-K 
detector, are compared to the total number and spectrum 
calculated { from} the results of the near detectors under the assumption 
of the absence of neutrino oscillations.
For the measurement of the energy of neutrinos in the S-K detector, 
quasi elastic one-ring events
$\nu_{\mu}+n\to \mu^{-}+p$ are selected. 

The first results of the K2K experiment were recently published~\cite{K2K}.
The total number of muon events observed in the S-K detector is equal to 56:
it must be compared with an expected number of  events equal to 
$80.1 ^{+6.2}_{-5.4}$. The observed number of the one-ring muon events 
{ which} was used for the calculation of neutrino spectrum is equal to 29,
{ while} the expected number of the one-ring events is equal to 44.

Thus, in the long baseline accelerator K2K experiment, indications in 
favour of the disappearance  of the accelerator $\nu_{\mu}$ were obtained.
From the  maximum likelihood two-neutrino analysis of the data  the following 
best-fit values of the oscillation parameters were found
\be
\sin^{2}2\,\theta_{\rm {K2K}}=1:\,~ \Delta m^{2}_{\rm{K2K}}= 2.8\,~10^{-3}\,~ \rm{eV}^{2}.
\label{096}
\ee
These values are in agreement with the values of the oscillation parameters 
 found from the analysis of the S-K atmospheric neutrino data 
(see (\ref{095})).
The first K2K results were obtained with $4.8\,~10^{19}$ protons on target 
(POT). It is expected that $10^{20}$ POT
will be utilized in the experiment.

\subsection{Evidence in favour of { the} transitions of solar $\nu_{e}$ 
into $\nu_{\mu}$ and $\nu_{\tau}$}

The energy of the sun is produced in the reactions of the thermonuclear
pp and CNO cycles, in which protons and electrons are converted into 
Helium and electron neutrinos: 

$$ 4p + 2 e^{-} \to ^{4}\rm{He}+ 2\nu_{e}.$$

The reactions, { which are} most important for the solar neutrino 
experiments, are listed in Table I. From this Table one can see that
the largest part of the solar neutrino flux draw up low 
energy $\rm{pp}$ neutrinos. 
According to the SSM BP00~\cite{BPin}, the medium energy mono-energetic 
$^{7}\rm{Be}$ neutrinos make up about 10 \% of the total flux, { while} 
the high energy $^{8}\rm{B}$ neutrinos constitute only about $10^{-2}$ \% 
of the total flux. However, the $^{8}\rm{B}$ decay is { a} very 
important source of the solar neutrinos: in the S-K~\cite{S-Ksol} and 
SNO~\cite{SNO,SNONC,SNOCC} experiments, due to { the} high energy 
thresholds, practically only neutrinos from $^{8}\rm{B}$-decay 
can be detected.\footnote{ According to the SSM BP00 the flux of the 
high energy $hep$ neutrinos, produced in the reaction 
$ ^{3}\rm{He}+ p \to ^{4}\rm{He} + e^{+} + \nu_{e}$,
is about three orders of magnitude smaller than the flux of the 
$^{8}\rm{B}$ neutrinos.}
{ The} $^{8}\rm{B}$ neutrinos { also} give { the} dominant 
contribution to the event rate measured in the Homestake experiment~\cite{Cl}.

\begin{center} 
 Table I\\
\vspace{0.2cm}
\begin{tabular}{|ccc|}
\hline
Reaction
&
Neutrino energy
&
SSM BP00 flux
\\
\hline
$p\, p \to d\,  e^{+}\, \nu_{e}$
&
$\leq 0.42\, \rm{MeV}$
&
$5.95\cdot 10^{10}\,\rm{cm}^{-2}\,\rm{s}^{-1}$
\\
$e^{-}+ ^{7}\rm{Be}\to \nu_{e}\,^{7}\rm{Li}$ 
&
$ 0.86\, \rm{MeV}$
&
$4.77\cdot 10^{9}\,\rm{cm}^{-2}\,\rm{s}^{-1}$
\\
$^{8}\rm{B} \to ^{8}\rm{Be}^{*}\,e^{+}\, \nu_{e}$
&
$\leq 15\, \rm{MeV}$
&
$5.05\cdot 10^{6}\,\rm{cm}^{-2}\,\rm{s}^{-1}$
\\
\hline
\end{tabular}
\end{center}
\begin{center} 
The main sources of the solar neutrinos. The maximum neutrino energies 
and SSM BP00~\cite{BPin} fluxes are also given.
\end{center}

The event rates measured in all solar neutrino experiments 
are significantly smaller than the 
{ ones} predicted by the Standard Solar models.

In the Homestake experiment~\cite{Cl} solar neutrinos are detected 
{ through} the observation of the Pontecorvo-Davis reaction
$\nu_{e}+  ^{37}\rm{Cl} \to e^{-}+  ^{37}\rm{Ar}$, in the GALLEX-GNO~\cite{GALLEX-GNO} and SAGE~\cite{SAGE} experiments 
{ through} the reaction  $\nu_{e}+  ^{71}\rm{Ga} \to e^{-}+  ^{71}\rm{Ge}$ 
and in the S-K~\cite{S-Ksol}  experiment solar neutrinos are detected via 
the observation of the process $\nu+  e \to \nu+  e$.

For the ratio R of the observed and the predicted by SSM BP00~\cite{BPin} 
rates the following values were obtained: 

\bea
 &&R = 0.34 \pm 0.03 \,~~~~ (\mathrm{Homestake}) 
\nonumber\\
&&R = 0.58 \pm 0.05 \,~~~~ (\mathrm{GALLEX-GNO})
\nonumber\\
 &&R= 0.60 \pm 0.05 \,~~~~ (\mathrm{SAGE}))
\nonumber\\
 &&R = 0.465 \pm 0.018  \,~~ (\mathrm{S-K})\,.
\nonumber
\eea

If there is neutrino mixing the original solar $\nu_{e}$'s,
due to neutrino oscillations or matter MSW transitions, are transfered into 
another type of neutrinos, which { can} not be detected in the radiochemical
Homestake, GALLEX-GNO and SAGE experiments. In the S-K experiment
mainly  $\nu_{e}$ are detected: the sensitivity of the
experiment to $\nu_{\mu}$ and $\nu_{\tau}$ is about six times smaller than the
sensitivity to $\nu_{e}$.
Thus, neutrino oscillations or MSW transition in matter
provide a natural explanation { for} the { observed} 
depletion of the fluxes of solar $\nu_{e}$.

Recently { a} strong model independent evidence in favour of the 
transition of the solar $\nu_{e}$ into  $\nu_{\mu}$ and $\nu_{\tau}$ 
was obtained in the SNO experiment~\cite{SNO,SNONC,SNOCC}.  
The detector in the SNO experiment  is a heavy water Cherenkov detector 
(1 kton of $\rm{D}_{2}O$).
Neutrinos from the Sun are detected via the observation of
the following three reactions:

\bea
&1.~~{\rm{CC~~ reaction}}~~~~~~~~~~&\nu_e + d \to e^{-}+ p +p\,,
\label{097}\\
&2.~~{\rm{NC~~ reaction}}~~~~~~~~~~&\nu + d \to \nu + n +p\,,
\label{098}\\
&3.~~{\rm{ES~~ process}}~~~~~~~~~~&\nu + e \to \nu + e \,
\label{099}
\eea
During 306.4 days of running  $1967^{+61.9}_{-60.9}$  
CC events, $576.5^{+49.5}_{-48.9}$ NC events,
and  $263.6^{+26.4}_{-25.6}$ 
ES events were recorded in the SNO experiment. 
The kinetic energy threshold for the detection of electrons { was} 
equal to 5~MeV, the NC threshold  to 2.2~MeV. 
Thus, practically only neutrinos from { the decay
$^{8}\rm{B}\to ^{8}\rm{Be}+e^{+}+\nu_{e} $ are detected in 
the SNO experiment. The important point is that the initial spectrum of $^{8}\rm{B}$
neutrinos is known~\cite{Ortiz}.}


The total CC event rate can be presented in the form

\be
R_{\nu_{e}}^{CC}=  <\sigma^{CC}_{\nu_{e}d}>\Phi_{\nu_{e}}^{CC}\,,
\label{100}
\ee
where $<\sigma^{CC}_{\nu_{e}d}>$ is the cross section of the CC 
process (\ref{097}), averaged over the initial spectrum of
$^{8}\rm{B}$ neutrinos, and $\Phi_{\nu_{e}}^{CC}$
is the flux of $\nu_e$ on the Earth, { which} is given by the relation
\be
\Phi_{\nu_{e}}^{CC} = <P(\nu_e \to\nu_e)>_{CC}\,~\Phi_{\nu_{e}}^{0}\,, 
\label{101}
\ee
where $\Phi_{\nu_{e}}^{0}$ is the total initial flux of  $\nu_e$
and $<P(\nu_e \to\nu_e)>_{CC}$ is the averaged $\nu_e$ survival probability.

All flavour neutrinos $\nu_e$, $\nu_{\mu}$ and $\nu_{\tau}$ are recorded via
the detection of the NC process (\ref{098}). Taking into account 
{ the} $\nu_{e}-\nu_{\mu}- \nu_{\tau}$ universality of the neutral 
current interaction, for the total NC event rate we have

\be
R^{NC}_{\nu}=  <\sigma^{NC}_{\nu d}>\Phi_{\nu}^{NC}\,,
\label{102}
\ee
where  $<\sigma^{NC}_{\nu d}>$ is the cross section of the NC process
(\ref{098}), averaged over the initial spectrum of the 
$^{8}\rm{B}$ neutrinos, and  $\Phi_{\nu}^{NC}$
is the total flux of all flavour neutrinos on the Earth.
We have 

\be
\Phi_{\nu}^{NC}=\sum_{l=e,\mu,\tau}\Phi_{\nu_l}^{NC}\,,
\label{103}
\ee
{ where }
\be
\Phi_{\nu_l}^{NC}=  <P(\nu_e \to\nu_l)>_{NC}\,~\Phi_{\nu_{e}}^{0}\,, 
\label{104}
\ee
$<P(\nu_e \to\nu_l)>_{NC}$ 
{ being} the averaged probability of the 
transition $\nu_e \to\nu_l$ ($l=e,\mu,\tau$).

All flavour neutrinos are detected also via the observation of the ES process
(\ref{099}). However, the cross section of the NC process
$\nu_{\mu,\tau}+e \to\nu_{\mu,\tau}+e$ is about six times smaller
than the cross section of the CC and NC process $\nu_{e}+e \to\nu_{e}+e$.

The total ES event rate can be presented in the form

\be
R^{ES}_{\nu}=  <\sigma_{\nu_{e} e}>\Phi_{\nu}^{ES}.
\label{105}
\ee
Here $<\sigma_{\nu_{e} e}>$ is the cross section of the process 
$\nu_{e}e \to\nu_{e}e$, 
averaged over initial spectrum of the $^{8}B$ neutrinos,
\be
\Phi_{\nu}^{ES}= \Phi_{\nu_{e}}^{ES} + \frac{<\sigma_{\nu_{\mu} e}>}
{<\sigma_{\nu_{e} e}>}\,~\Phi_{\nu_{\mu,\tau}}^{ES}\,,
\label{106}
\ee
where 
$\Phi_{\nu_{e}}^{ES}$ ( $\Phi_{\nu_{\mu,\tau}}^{ES}$) is 
the flux of $\nu_{e}$ ($\nu_{\mu}$ and $\nu_{\tau}$) and
\be
\frac{<\sigma_{\nu_{\mu} e}>}
{<\sigma_{\nu_{e} e}>}
\simeq 0.154.
\label{107}
\ee
We have
\be
\Phi_{\nu_{l}}^{ES}= <P(\nu_e \to\nu_l)>_{ES}\,~\Phi_{\nu_{e}}^{0},
\label{108}
\ee
where
$<P(\nu_e \to\nu_l)>_{ES}$ is the averaged probability of the 
transition $\nu_e \to\nu_l$.

In the SNO experiment it was found \cite{SNOCC}
\be
(\Phi_{\nu}^{ES})_{\rm{SNO}} =({2.39^{+0.24}_{-0.23}~ \mbox{(stat.)}
\pm 0.12~\mbox{(syst.)}} ) \cdot 10^{6}\,~
\rm{cm}^{-2}\rm{s}^{-1} \,,
\label{109}
\ee
This value is in a good agreement with 
the value of the ES flux, obtained in the Super-Kamiokande experiment.

In the S-K solar neutrino experiment~\cite{S-Ksol} neutrinos are detected 
via the observation of the ES process.
During 1496 days of running a large number $22400 \pm 800$
of solar neutrino events with recoil total energy threshold { of} 5 MeV
were recorded. From the data of the S-K experiment the value 
\be
(\Phi_{\nu}^{ES})_{\rm{S-K}} =({2.35 \pm 0.02~ \mbox{(stat.)} 
\pm0.08~\mbox{(syst.)}} ) \cdot 10^{6}\,~\rm{cm}^{-2}\rm{s}^{-1} \,.
\label{110}
\ee
{  was obtained}.

 In the S-K experiment the spectrum of the recoil electrons was measured: 
no sizable distortion { was observed} with respect 
to the spectrum, expected under the assumption of no neutrino oscillations.

The spectrum of electrons, 
produced in the CC process (\ref{097}), was  measured in the
SNO experiment~\cite{SNOCC}. Also in
this experiment no distortion of the electron spectrum was observed.

Thus, the data of the S-K and SNO experiments are compatible with the 
assumption that in the high-energy $^{8} B$ region the probability of 
solar neutrinos to survive is practically a constant:
\be
P(\nu_e \to\nu_e) \simeq \rm{const}.
\label{111}
\ee
The constant $\nu_{e}$ survival probability, { Eq.~}(\ref{111}), 
implies
$$<P(\nu_e \to\nu_e)>_{CC}\simeq <P(\nu_e \to\nu_e)>_{NC}
\simeq <P(\nu_e \to\nu_e)>_{ES}\,.$$
Taking into account these relations, 
from (\ref{101}), (\ref{104}) and (\ref{108}) 
in the high energy $^{8} B$ region we obtain
\be
\Phi_{\nu_{e}}^{CC}\simeq \Phi_{\nu_{e}}^{NC}\simeq
\Phi_{\nu_{e}}^{ES}\,.
\label{112}
\ee

In the SNO experiment for the flux of $\nu_e$ on the Earth it was found 

\be
(\Phi_{\nu_{e}}^{CC})_{\rm{SNO}} 
=({1.76^{+0.06}_{-0.05}\mbox{(stat.)}^{+0.09}_{-0.09}~\mbox{(syst.)}} ) 
\cdot 10^{6}\,~\rm{cm}^{-2}\rm{s}^{-1}.
\label{113}
\ee
For the flux of all flavour neutrinos $\Phi_{\nu}^{NC}$ from the NC measurement
it was obtained the value
\be
(\Phi_{\nu}^{NC})_{\rm{SNO}} =({5.09^{+0.44}_{-0.43}
\mbox{(stat.)}^{+0.46}_{-0.43}~\mbox{(syst.)}} ) \cdot 10^{6}\,~
\rm{cm}^{-2}\rm{s}^{-1}\,, 
\label{114}
\ee
which is about three times larger than the flux of 
electron neutrinos on the Earth.

\begin{figure}[t]
\begin{center}
\includegraphics*[bb=5 0 521 390, height=7cm]{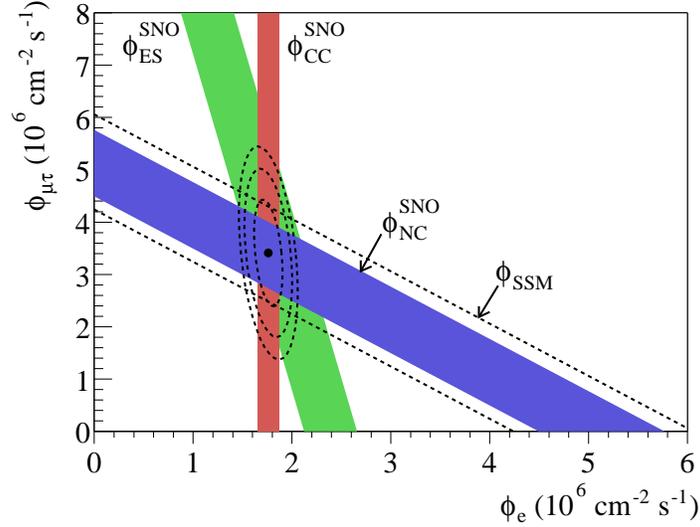}
\end{center}
\caption{ \label{sno-0204008-fig3}
Flux of $\nu_\mu$ and $\nu_\tau$ vs flux of $\nu_e$
in the ${}^{8}$B energy region
deduced from the three neutrino reactions in SNO.
The diagonal bands show the total ${}^{8}$B flux
as predicted by the BP00 SSM~\cite{BPin} (dashed lines)
and that measured in SNO experiment via the observation of
NC (solid band).  The interception of these bands with the 
axes represent the $\pm 1\sigma$ errors. The
bands intersect at the fit values for
$\Phi_{\nu_e}$ and $\Phi_{\nu_{\mu,\tau}}$.
Figure taken from Ref.~\cite{SNONC}.
}
\end{figure}

{ Obviously} the NC flux is given by
\be
\Phi_{\nu}^{NC}= \Phi_{\nu_e}^{NC}+ \Phi_{\nu_{\mu,\tau}}^{NC}\,
\label{115}
\ee
where $\Phi_{\nu_e}^{NC}$ is the flux of $\nu_{e}$ and
$\Phi_{\nu_{\mu,\tau}}^{NC}$ is the flux of $\nu_{\mu}$ and 
$\nu_{\tau}$.

{ By} combining now CC and NC fluxes and using the relation (\ref{112}), 
we can determine  the flux $\Phi_{\nu_{\mu,\tau}}^{NC}$.
In {Refs.}~\cite{SNOCC,SNONC} the ES flux (\ref{109}) was also taken
into account(see Fig.~\ref{sno-0204008-fig3}).
For the flux of $\nu_{\mu}$ and $\nu_{\tau}$ on the 
Earth the following value 
\be
(\Phi_{\nu_{\mu,\tau}})_{SNO} =(3.41^{+0.45}_{-0.45}
\mbox{(stat.)}^{+0.48}_{-0.45}~\mbox{(syst.)})\cdot 10^{6}\,~
\rm{cm}^{-2}\rm{s}^{-1} 
\label{116}
\ee
was { then} obtained. Thus, { the} detection of solar neutrinos   
via the  simultaneous observation of CC, NC and ES processes 
allowed the SNO collaboration to obtain {\em a direct model independent 
$5.3\,~ \sigma$ evidence of the presence of $\nu_{\mu}$ and $\nu_{\tau}$ 
in the flux of the solar neutrinos on the Earth.}

The total flux of the $^{8} B$ neutrinos, predicted by SSM BP00~\cite{BPin},
is equal to
\be
(\Phi_{\nu_{e}}^{0})_{\rm{SSM\, BP}}= (5.05^{+1.01}_{-0.81})\cdot 10^{6}\,~
\rm{cm}^{-2}\rm{s}^{-1}\,.  
\label{117}
\ee
This value is compatible with the value of total flux of all flavour neutrinos
(\ref{114}), determined from the data of the SNO experiment.

The flux of $\nu_{\mu}$ and $\nu_{\tau}$ on the Earth can be also obtained
from the SNO CC data and the S-K ES data. In the first SNO 
publication~\cite{SNO}  the value
\be
(\Phi_{\nu_{\mu,\tau}})_{\rm{S-K, SNO}} =(3.69 \pm 1.13 )\cdot 10^{6}\,~
\rm{cm}^{-2}\rm{s}^{-1} \,,
\label{118}
\ee
{ was found}, which is in a good agreement with the value (\ref{116}).

The data of all solar neutrino experiments can be described if { one 
assumes} that there are transitions of  solar $\nu_e$ into 
$\nu_{\mu}$ and $\nu_{\tau}$ and { that the} $\nu_e$ survival 
probability is given by the two-neutrino expression,
which is characterized by two oscillation parameters,  
$\Delta m^{2}_{\rm{sol}}$ and $\tan^{2}\theta_{\rm{sol}}$.
From the global $\chi^{2}$ fit of the total event rates, measured in
all solar neutrino experiments, several allowed regions (solutions)
in the plane of 
the oscillation parameters were obtained (see, for example, 
{ Ref.}~\cite{BGP}): LMA, LOW, SMA,
VO and other regions. The situation changed after
the recoil electron spectrum was measured in the S-K experiment~\cite{S-Ksol} 
and { the} SNO results~\cite{SNO,SNOCC,SNONC} were obtained. 
Analyzes  of all solar neutrino data showed that 
the large mixing angle LMA MSW region is the most plausible 
one (see \cite{Lisi} and references therein). In { Ref.}~\cite{SNOCC}  
the following best-fit LMA values of the solar neutrino oscillation parameters 
were found:

\be
\Delta m^{2}_{\rm{sol}}=5\cdot 10^{-5}\rm{eV}^{2};
\,\tan^{2}\theta_{\rm{sol}}=0.34;\,~
\chi^{2}_{\rm{min}}= 57/72\, \rm{d.o.f.}.
\label{119}
\ee

Recent data of the KamLAND experiment~\cite{Kamland} (see below) allow
to exclude all solutions of the solar neutrino problem, except the LMA one.

\subsection{Reactor experiments CHOOZ and Palo Verde}

\begin{figure}[t]
\begin{center}
\includegraphics*[bb=0 0 434 360, height=7cm]{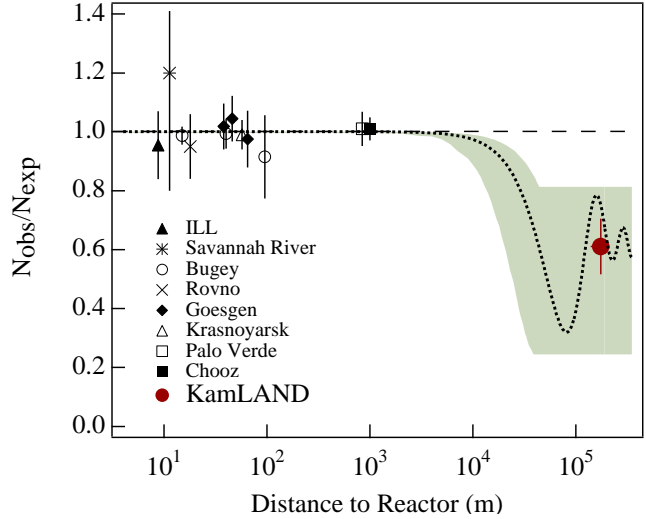}
\end{center}
\caption{ \label{kamland-0212021-1}
The ratio of measured to expected numbers of $\bar \nu_e$ events for 
different  reactor neutrino experiments.
The shaded region indicates the range of the 
predictions corresponding to the 95\% C.L. LMA region found from the 
global analysis of the solar neutrino data~\cite{Fogli}.
The dotted curve corresponds to the best-fit values
$\Delta{m}^{2}_{\mathrm{sol}} = 5.5 \times 10^{-5} \, \mathrm{eV}^{2}$
and $\sin^{2}2\vartheta_{\mathrm{sol}} = 0.83$
found in Ref.~\cite{Fogli}. Figure taken from Ref.~\cite{Kamland}.
}
\end{figure}

In the long baseline reactor experiments CHOOZ~\cite{CHOOZ} and 
Palo Verde~\cite{PaloV} the disappearance of the reactor $\bar\nu_{e}$'s
in the atmospheric range of $\Delta m^{2}$ { was} searched for.
In spite of { the fact that} in these experiments no 
indication in favour of neutrino oscillations were found,
 their results are very important for the neutrino mixing.

In the CHOOZ experiment $\bar\nu_{e}$'s from two reactors at { a} 
distance of about 1 km from the detector were recorded via the observation 
of the process
$$\bar\nu_{e}+ p \to e^{+}+ n. $$ 
The { value of the} ratio $R$ of the total number of the detected 
$\bar\nu_{e}$ events to the expected number  was found { to be}
$$ R =1.01 \pm 2.8 \%\,(\rm{stat})\pm\pm 2.7 \%\,(\rm{syst})\,~~( \rm{CHOOZ})$$
In the similar Palo Verde experiment it was obtained:
$$R =1.01 \pm 2.4 \%\,(\rm{stat})\pm 5.3 \%\,(\rm{syst})
\,~~(\rm{Palo Verde}) $$

The data of the experiments were analyzed in~\cite{CHOOZ,PaloV} in the 
framework of two-neutrino oscillations and exclusion plots in the plane 
of the oscillation parameters $\Delta m^{2}$ and $\sin^{2}2\,\theta$ were 
obtained. From the CHOOZ exclusion plot at $\Delta m^{2}=2.5 \cdot 10^{-3}
\,~\rm{eV}^{2}$ (the S-K best-fit value) { one gets}
$$(\sin^{2}2\,\theta)_{\rm{CHOOZ}} \lesssim 1.5\cdot 10^{-1}.$$

\begin{figure}[t]
\begin{center}
\includegraphics*[bb=69 50 520 494, height=7cm]
{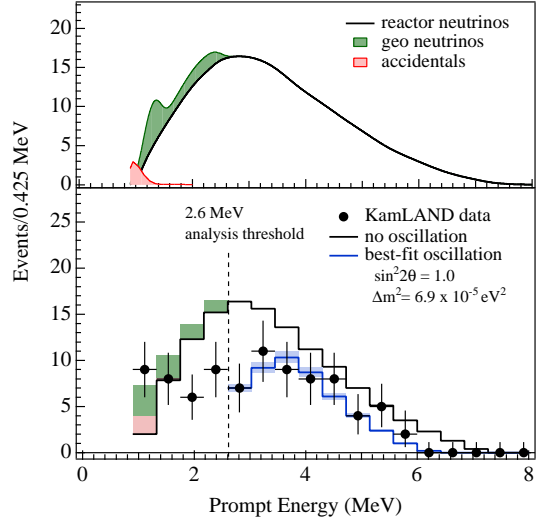}
\end{center}
\caption{ \label{kamland-0212021-2}
Upper panel: Expected reactor $\bar\nu_e$ energy spectrum with contributions
of $\bar\nu_{\mathrm{geo}}$ (antineutrinos produced in $^{238}$U and 
$^{232}$Th decays in the Earth) and accidental background.
Lower panel: Energy spectrum of the observed prompt events (solid 
circles with error bars), along with the expected no oscillation 
spectrum (upper histogram, with $\bar\nu_{\mathrm{geo}}$ and accidentals 
shown) and best fit (lower histogram) in the case of neutrino oscillations. 
The shaded band indicates the systematic error in the best-fit spectrum.
The vertical dashed line corresponds to the threshold { of} 2.6~MeV.
Figure is taken from Ref.~\cite{Kamland}.
}
\end{figure}

\subsection{The KamLAND evidence in favour of the disappearance of the 
reactor $\bar\nu_{e}$}

The first results of the KamLAND experiment~\cite{Kamland} were published
recently. In this experiment  $\bar\nu_{e}$'s  from many reactors in Japan 
and Korea are detected via the observation of the classical process
$$\bar\nu_{e}+p \to e^{+}+n $$
The threshold of this process { is} $\simeq$ 1.8 MeV.
About 80 \% of the total number of the events is due to $\bar\nu_{e}$ 
from 26 reactors within the distances of 138-214 km.
\begin{figure}[h]
\begin{center}
\includegraphics*[bb=0 0 567 539, height=7cm]{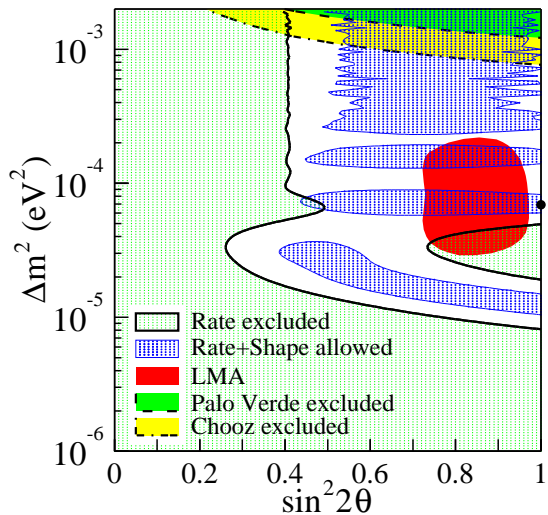}
\end{center}
\caption{ \label{kamland-0212021-3}
KamLAND excluded regions of neutrino oscillation parameters
$\Delta m^{2}$ and $\sin^{2}2\theta $
for the rate analysis and allowed regions for the combined rate and
energy spectrum analysis at 95\% C.L. At the top are the 95\% C.L. 
excluded region from CHOOZ~\cite{CHOOZ} and Palo Verde~\cite{PaloV} 
experiments, respectively. The dark area is the 95\% C.L. LMA allowed
region obtained in Ref.~\cite{Fogli}. The thick dot indicates the 
best-fit values of the oscillation parameters.
Figure taken from Ref.~\cite{Kamland}.
}
\end{figure}

The 1 kt liquid scintillator detector of the  KamLAND experiment is 
located in the Kamioka mine at { a} depth of about 1 km.
Both prompt photons from the annihilation of $e^{+}$ in the scintillator 
and 2.2~MeV delayed photons from the neutron capture $n+p\to d +\gamma$  
are detected. The mean neutron capture time is $188\pm 23\, \mu$sec.
In order to avoid background, mainly from the the decay of
$^{238}\rm{U}$ and $^{232}\rm{Th}$ in the Earth, the cut 
$E_{prompt}>2.6\, \rm{MeV}$ was applied.

During 145.1 days of running { there were} observed 54 $\bar\nu_{e}$ 
events. The number of the events expected in the { absence of} neutrino 
oscillations is equal to $86.8 \pm 5.6$.
For the ratio of observed and expected 
$\bar\nu_{e}$ events the following value
\be
\frac{N_{obs}-N_{BG}}{N_{exp}} = 0.611 \pm 0.085 \pm 0.041.
\label{120}
\ee
was obtained.

In Fig.~\ref{kamland-0212021-1}, for all reactor neutrino experiments, 
the dependence of the ratio of the observed and expected 
$\bar\nu_{e}$ events on the average distance between reactors
and detectors is plotted. 
The dotted curve was calculated with the best-fit solar neutrino LMA values
of the oscillation parameters $\Delta m^{2} = 5.5 \cdot 10^{-5}\rm{eV}^{2}$
and $\sin^{2}2\,\theta =0.83$, obtained in Ref. \cite{Fogli}.

In the KamLAND experiment the prompt energy spectrum 
was also measured (see Fig.~\ref{kamland-0212021-2}).
 The prompt energy is connected with
the energy of $\bar\nu_{e}$ by the relation $E_{prompt}= E_{\bar\nu_{e}}-
0.8\, \rm{ MeV} - \bar{E_{n}}$ ($\bar{E_{n}}$ { being} the average 
energy of the neutron). From the two-neutrino analysis of the KamLAND data 
the following best-fit values of the oscillation parameters were obtained
$$(\Delta m^{2})_{\rm{KamLAND}} = 6.9 \cdot 10^{-5}\rm{eV}^{2};\,~
(\sin^{2}2\,\theta)_{\rm{KamLAND}} =1\,.$$

These values are compatible with the values  of the oscillation 
parameters in the solar neutrino LMA MSW region. 
The allowed region in the plane of the oscillation parameters, obtained from 
{ the} analysis of the measured rate and measured spectrum is shown in
Fig.~\ref{kamland-0212021-3} (95 \% CL). The region outside the solid line 
is excluded from the rate analysis. The dark region is the solar neutrino 
LMA allowed region, obtained in Ref.~\cite{Fogli}.

The KamLAND results provide { a} strong evidence for neutrino masses
and oscillations, obtained for the first time in an experiment 
with terrestrial antineutrinos with the expected flux well under control.
It allows us to exclude { the} SMA, LOW and VAC regions of neutrino 
oscillation parameters. The only viable solution of the solar neutrino 
problem { appears to be} the LMA MSW solution.

\section{Neutrino oscillations in the framework of three-neutrino mixing}

\subsection{Neutrino oscillations in the atmospheric range of 
the neutrino mass-squared difference} 

We have discussed evidences in favour of neutrino oscillations, { which} 
were obtained in the solar, atmospheric and reactor KamLAND neutrino 
experiments. At present { there} exists also an indication in favour of 
the transition $\bar \nu_{\mu} \to \bar \nu_{e}$, that was obtained in 
the single accelerator LSND experiment~\cite{LSND}. The LSND data can 
be explained by neutrino oscillations. From { the} 
analysis of the data, { the following ranges for the 
values of the oscillation parameters}
\bea
&& 2\cdot 10^{-1}\lesssim (\Delta m^{2})_{\rm{LSND}}\lesssim 1\,\rm{eV}^{2};
\nonumber\\
&& 3\cdot 10^{-3}\lesssim (\sin^{2}2\,\theta)_{\rm{LSND}}
\lesssim 4\cdot 10^{-2}
\nonumber
\eea
were obtained.

In order to describe the data of the solar, atmospheric, KamLAND {\em and} 
LSND experiments, which require three different values of neutrino 
mass-squared differences, it is necessary to assume that there are (at least) 
four massive and mixed neutrinos. This means that in addition to { the} 
three flavour neutrinos (at least) one sterile neutrino must exist 
(see, for example, { Ref.}~\cite{BGG}).
The result of the LSND experiment requires, however, confirmation. { The}
MiniBooNE experiment at Fermilab~\cite{MiniB}, { which} started in 2002,
is aimed at checking the LSND result.

We will consider here the {\em minimal} scheme of three-neutrino mixing

\be
\nu_{\alpha L}=\sum_{i=1}^{3}\,U_{\alpha i}\,\nu_{i L},
\label{121}
\ee
where $U$ is the unitary 3$\times$3 PMNS mixing matrix. This scheme
provides two independent $\Delta m^{2}$'s and allows us to describe solar, 
atmospheric, KamLAND and other neutrino oscillation data.

Let us start with the consideration of neutrino oscillations in the 
atmospheric range of $\Delta m^{2}$, which can be explored in the 
atmospheric and long baseline accelerator and reactor neutrino experiments.
In the framework of the three-neutrino mixing (\ref{121}),
with $m_1 < m_2< m_3$, there are two possibilities:
\begin{description}
\item{I.} Hierarchy of neutrino mass-squared differences
\be
\Delta m^{2}_{21}\simeq \Delta m^{2}_{\rm{sol}};
\,
\Delta m^{2}_{32}\simeq \Delta m^{2}_{\rm{atm}};
\,~\Delta m^{2}_{21} \ll \Delta m^{2}_{32}.
\label{122}
\ee
\item{II.} Inverted hierarchy of neutrino mass-squared differences
\be
\Delta m^{2}_{32}\simeq \Delta m^{2}_{\rm{sol}};
\,
\Delta m^{2}_{21}\simeq \Delta m^{2}_{\rm{atm}};
\,~\Delta m^{2}_{32} \ll \Delta m^{2}_{21}.
\label{123}
\ee
\end{description}

We will first assume that { the} neutrino mass spectrum is of the type I.
The values of  ${L}/{E}$, relevant  for neutrino oscillations in 
the atmospheric range of neutrino mass-squared difference,
satisfy the inequality:

$$\Delta m^{2}_{21}\,~ \frac{L}{E}\ll 1.$$
Thus, we can neglect the contribution of $\Delta m^{2}_{21}$ to 
the transition probability, Eq.~(\ref{034}). { In this case,} for the
probability of the transition $\nu_\alpha \to \nu_{\alpha'}$
we obtain  the following expression
\be
{\mathrm P}(\nu_\alpha \to \nu_{\alpha'}) \simeq
\left|\delta_{{\alpha'}\alpha} + U_{\alpha' 3}  U_{\alpha 3}^*
\,~ (e^{- i \Delta m^2_{32} \frac {L} {2E}} -1)\right|^2 \,.
\label{124}
\ee
Hence, in the leading approximation, { the} transition probabilities
in the atmospheric range of $\Delta m^2$ are determined by the largest 
neutrino mass-squared difference $\Delta m^2_{32}$ and { by} the elements  
of the third column of the neutrino mixing matrix, which connect { the}
flavour neutrino fields $\nu_{\alpha L}$ with the field of the heaviest 
neutrino $\nu_{3L}$.

For the appearance probability  we obtain from { Eq.}~(\ref{124}) 
the expression
\begin{equation}
{\mathrm P}(\nu_\alpha \to \nu_{\alpha'}) =
 \frac{1}{2} {\mathrm A}_{{\alpha'};\alpha}
\left(1 - \cos \Delta m^{2}_{32}\, \frac {L} {2E}\right)
\,~(\alpha \not= \alpha'),
\label{125}
\end{equation}
where the oscillation amplitude is given by 
\be
{\mathrm A}_{{\alpha'};\alpha}= 4\,~|U_{\alpha' 3}|^{2}\,~|U_{\alpha 3}|^{2}\,.
\label{126}
\ee

The survival probability can be obtained from 
the condition of conservation of probability and from Eqs.~(\ref{125}) and
~(\ref{125} we  have
\be
{\mathrm P}(\nu_\alpha \to \nu_\alpha)= 1 -\sum_{\alpha' \not= \alpha}\,
 {\mathrm P}(\nu_\alpha \to \nu_{\alpha'}) =1 - \frac{1}{2}
{\mathrm B}_{\alpha ; \alpha}
\left(1 - \cos \Delta m^{2}_{32}\, \frac {L} {2E}\right)\,.
\label{127}
\ee
Taking into account the unitarity of the mixing matrix, { the}
amplitude ${\mathrm B}_{\alpha;\alpha}$ can be written as: 
\be
{\mathrm B}_{\alpha; \alpha}\equiv \sum_{\alpha' \not= \alpha}\,
{\mathrm A}_{{\alpha'};\alpha}=
4\,|U_{\alpha 3}|^{2}\,\left(1 -|U_{\alpha 3}|^{2}\right)\,.
\label{128}
\ee
{ Obviously,}  in the case of  inverted hierarchy of the 
neutrino mass-squared differences (case II above) { the} 
transition probabilities can be obtained from 
{ Eqs.}~(\ref{125})-(\ref{128}) { with the replacements}
$\Delta m^{2}_{32}\to \Delta m^{2}_{21}$ and $|U_{\alpha 3}|^{2}\to
|U_{\alpha 1}|^{2}$.

{ We notice that the} transition probability (\ref{125}) depends only on 
$|U_{\alpha 3}|^{2}$ and $\Delta m^{2}_{32}$. The CP phase does not
enter into this equation and in the leading approximation the 
relation
\be
{\mathrm P}(\nu_\alpha \to \nu_{\alpha'})=
{\mathrm P}(\bar \nu_\alpha \to \bar \nu_{\alpha'})
\label{129}
\ee
is automatically satisfied.

Thus, { the} investigation of { the} effects of CP violation in 
the lepton sector in the future long baseline neutrino oscillation 
experiments will be a difficult problem: possible effects are suppressed 
due to the smallness of the parameter 
${\Delta m^{2}_{12}}/{\Delta m^{2}_{32}}$.
High precision experiments on the search for effects 
of  CP-violation in the lepton sector are planned for the future neutrino 
facility JHF~\cite{Nakaya} and Neutrino Factories~\cite{Lindner,Dydak}.

{ The} transition probabilities (\ref{125}) and (\ref{127}) 
have a {\em two-neutrino form} in { each} channel.
This is the obvious consequence of the fact that only the largest 
mass-squared difference $\Delta{m}^2_{32} $ contributes to the transition 
probabilities. The elements  $|U_{\alpha 3}|^{2}$, which determine the 
oscillation amplitudes, satisfy the unitarity condition 
$\sum_{\alpha}|U_{\alpha 3}|^{2}=1 $. Hence, in leading approximation, 
transition probabilities are characterized by three parameters.
We can choose the { latter to be}:
$$\Delta m^{2}_{32},\,\,\,~\sin^{2}\theta_{23},\,\,\,~|U_{e3}|^{2}$$ 
{ Then,} from Eqs.~(\ref{045}) and (\ref{126}), for the amplitudes of
$\nu_{\mu}\to\nu_{\tau}$ and $\nu_{\mu}\to\nu_{e}$ transitions  we obtain
the expressions
\bea
&& A_{\tau;\mu}= (1-|U_{e3}|^{2})^{2}\sin^{2}2\,\theta_{23};\,~
\nonumber\\
&& A_{e;\mu}=4\,|U_{e3}|^{2}\,(1-|U_{e3}|^{2}\sin^{2}\,\theta_{23}\,.
\nonumber
\eea

\subsection{Oscillations in the solar range of 
neutrino mass-squared difference}

Let us consider now, in the framework of the tree-neutrino mixing, 
neutrino oscillations in the solar range of $\Delta m^{2}$.
The $\nu_{e}$ survival probability in vacuum can be written in the form

\be
{\mathrm P}(\nu_e\to\nu_e)=
 \left|
\sum_{i=1, 2}| U_{e i}|^2 \,  e^{ - i \, 
\Delta{m}^2_{i1} \frac {L}{2 E} }
 + | U_{e 3}|^2  \,  e^{ - i \, 
\Delta{m}^2_{31} \frac {L}{2 E} }\,\right|^2
\label{130}
\ee
We are interested in the $\nu_{e}$ survival probability averaged over
the region where neutrinos are produced, over energy resolution etc. 
Because of the hierarchy $\Delta m^{2}_{32}\gg \Delta m^{2}_{21}$, 
in the expression for the averaged survival probability
the interference between the first and the second terms in 
{ Eq.}~(\ref{130}) disappears. The averaged survival probability can 
then be presented in the form
\be
{\mathrm P}(\nu_{e}\to\nu_{e})=|U_{e 3}|^{4}+ 
\left(1-|U_{e 3}|^{2}\right)^{2}\,~
P^{(1,2)}(\nu_{e}\to\nu_{e})\,.
\label{131}
\ee
Here $P^{(1,2)}(\nu_{e}\to\nu_{e})$ is given by the expression

\begin{equation}
{\mathrm P}^{(1,2)}(\nu_e \to \nu_e) =
 1 - \frac {1} {2}\,~A^{(1,2)}\,~ 
\left(1 - \cos \Delta m^{2}_{21} \frac {L} {2E}\right)\,,
\label{132}
\end{equation}
where
\be
A^{(1,2)}= 4\,\frac{ |U_{e 1}|^{2}\,|U_{e 2}|^{2}} 
{( 1-   |U_{e 3}|^{2})^{2} }.
\label{133}
\ee

With the help of Eq.~(\ref{037}), Eq.~(\ref{133}) simplifies to
\be
A^{(1,2)}=\sin^{2}2\,\theta_{12}\,.
\label{134}
\ee
Thus, the probability ${\mathrm P}^{(1,2)}(\nu_e \to \nu_e)$
is { again} characterized by two parameters { only} 
and has the standard two-neutrino form.

The expression (\ref{131}) is also valid for oscillations in
matter (see \cite{Schramm,BGG}). In this case
$P^{(1,2)}(\nu_{e}\to\nu_{e})$ is the two-neutrino $\nu_{e}$ survival
probability in matter, calculated under the condition that the
density of electrons $\rho_{e}(x)$ in the effective
Hamiltonian of the interaction of neutrino with matter is
changed by $(1-|U_{e 3}|^{2})\,\rho_{e}(x)$.

Hence, the $\nu_{e}$ survival probability is characterized, in the 
solar range of $\Delta m^{2}$, by three parameters
$$\Delta m^{2}_{21},\,\,\,~\tan^{2}\theta_{12},\,\,\,~|U_{e3}|^{2}.$$ 
The only common parameter for the atmospheric and solar ranges of 
$\Delta m^{2}$ is $|U_{e3}|^{2}$.
As we will see in the next subsection, from the data of
the reactor CHOOZ and Palo Verde experiments, this parameter 
{ turns out to be} small.

\subsection{The upper bound of $|U_{e 3}|^{2}$ from the CHOOZ data.
Decoupling of oscillations in the solar and atmospheric ranges of 
$\Delta m^{2}$}

The long baseline reactor experiments CHOOZ~\cite{CHOOZ} and Palo 
Verde~\cite{PaloV} are sensitive to the atmospheric range of 
$\Delta m^{2}$. No indications in favour of the disappearance of reactor 
$\bar\nu_e$  was obtained in these experiments.
From the analysis of { their} data 
{ a} stringent bound on the parameter $|U_{e 3}|^{2}$ was obtained,
as it is explained below.

In the framework of the three-neutrino mixing the probability of 
 $\bar\nu_e$ to survive in the atmospheric range of $\Delta m^{2}$ is given 
by the expression

\be
{\mathrm P}(\bar \nu_e \to \bar \nu_e) =1 - \frac {1} {2}
{\mathrm B}_{e; e}\,~ 
\left(1 - \cos \Delta m^{2}_{32} \frac {L} {2E}\right)\,,
\label{135}
\ee
where 
\be
{\mathrm B}_{e ; e} =4\, |U_{e 3}|^{2}\,\left(1-|U_{e 3}|^{2}\right)\,.
\label{136}
\ee
In { Refs.}~\cite{CHOOZ,PaloV} exclusion plots in the plane of the parameters
$\Delta m^{2}\equiv \Delta m^{2}_{32}$ and 
$\sin^{2}2\theta \equiv {\mathrm B}_{e; e}$
were obtained, { from which one has}
\be
{\mathrm B}_{e ; e} \leq {\mathrm B}_{e ; e}^{0},
\label{137}
\ee
where the upper bound ${\mathrm B}_{e ; e}^{0}$  depends on 
$\Delta m^{2}_{32}$. For the S-K allowed values of the parameter $\Delta m^{2}_{32}$,
from the CHOOZ exclusion plot we have
\be
1\cdot 10^{-1}\leq {\mathrm B}_{e ; e}^{0}\leq 2.4\cdot 10^{-1}.
\label{138}
\ee

From { Eqs.}~(\ref{136}) and (\ref{138}) { the following bounds
on} the parameter $|U_{e 3}|^{2}$ { can be easily obtained}:
\be
|U_{e 3}|^{2} \leq
\frac{1}{2}\,\left(1 - \sqrt{1- {\mathrm B}_{e ; e}^{0} }\right)\lesssim
\frac{1}{4}\,  {\mathrm B}_{e ; e}^{0}
\label{139}
\ee
or
\begin{equation}
|U_{e 3}|^{2} \gtrsim
\frac{1}{2}\,\left(1 + \sqrt{1- {\mathrm B}_{e ; e}^{0}}\right)\geq
1-\frac{1}{4}\, {\mathrm B}_{e ; e}^{0}\,.
\label{140}
\end{equation}
 
Thus, { the} parameter $|U_{e 3}|^{2}$ can be { either} small or large 
(close to one). This last possibility is excluded, however, by the
solar neutrino data. In fact, if  $|U_{e 3}|^{2}$ is large, from 
Eq.~(\ref{131}) it follows that in the whole range of the solar neutrino 
energies the probability of $\nu_{e}$ to survive is close to one 
in obvious contradiction with the solar neutrino data.
Hence, the upper bound of the parameter $|U_{e 3}|^{2}$ is given by
Eq.~(\ref{139}). At the S-K best-fit value 
$\Delta m^{2}_{32}=2.5\cdot 10^{-3}\rm{ eV}^{2}$
we { get}:
\be
|U_{e 3}|^{2}\leq 4\cdot 10^{-2}\,~ (95 \%\,~\rm{CL}).
\label{141}
\ee

Taking into account the accuracies of the present-day experiments,
one can neglect $|U_{e 3}|^{2}$ in the expressions for the transition 
probabilities. In this approximation, neutrino oscillations in the 
atmospheric range of $\Delta m^{2}$ are $\nu_{\mu}\to\nu_{\tau}$ 
oscillations, with $\Delta m^{2}_{32}\simeq \Delta m^{2}_{\rm{atm}}$ and
 $A_{\tau;\mu}\simeq \sin^{2}2\,\theta_{23}\simeq  
\sin^{2}2\,\theta_{\rm{atm}}$.
In the solar range of $\Delta m^{2}$ we have
\be
{\mathrm P}(\nu_e \to \nu_e)\simeq {\mathrm P}^{(1,2)}(\nu_e \to \nu_e),
\label{142}
\ee
where { the} two-neutrino survival probability
${\mathrm P}(\nu_e \to \nu_e)$ depends on the parameters 
$\Delta m^{2}_{21}\simeq \Delta m^{2}_{\rm{sol}}$ and
$\tan^{2}2\,\theta_{12}\simeq  \tan^{2}2\,\theta_{\rm{sol}}$.

Thus, due to the smallness of the parameter $|U_{e 3}|^{2}$ and  
{ the} hierarchy of neutrino mass squared differences 
$\Delta m^{2}_{12}\ll\Delta m^{2}_{32}$ neutrino
oscillations in the atmospheric and solar ranges of $\Delta m^{2}$, 
in the leading approximation, are decoupled~\cite{BG,BGG} and are 
described by { the} two-neutrino formulas, which are characterized, 
respectively, by the oscillation parameters 
$$\Delta m^{2}_{32},\,~~ \sin^{2}2\,\theta_{23}\,\,~~~
\rm{and}\,\,~~~
\Delta m^{2}_{21},\,~~ \tan^{2}\theta_{12}$$.
From the CHOOZ data only the upper bound of the parameter $|U_{e 3}|^{2}$
can be obtained. The possibilities to investigate effects of the three-neutrino
mixing and in particular important effects of CP violation in the lepton sector
depend on the value of this parameter. The value of the parameter
$|U_{e 3}|^{2}$ will be probed in MINOS
\cite{MINOS} and ICARUS \cite{ICARUS} experiments and 
in neutrino experiments at
  JHF \cite{Itow}
and Neutrino Factories (see \cite{Lindner,Dydak}).

From the data of the experiments on the investigation of neutrino oscillations
only neutrino mass-squared differences  
$\Delta m^{2}_{21}$ and  
$\Delta m^{2}_{32}$ can be determined. Neutrino masses $m_2$ and $m_3$ 
are given by the relations
\be
m_{2}= \sqrt{m^{2}_{1}+ \Delta m^{2}_{21}};\,~
m_{3}= \sqrt{m^{2}_{1}+ \Delta m^{2}_{32}+\Delta m^{2}_{21}}\,.
\label{143}
\ee

In the next section we will discuss the results of experiments which
allow us to obtain information about the absolute values of neutrino masses.

\section{$\beta$-decay experiments on the measurement of neutrino mass}

The standard method for the  measurement of the absolute value of { the}
neutrino {mass} is based on the detailed investigation of the 
high-energy part of the $\beta$-spectrum of the decay of tritium
\be
^{3}H \to ^{3}He + e^{-}+ \bar \nu_{e}\,.
\label{144}
\ee
This decay
is the
super-allowed one. Thus, the nuclear matrix element is a constant and 
{ the} electron spectrum is determined by the phase space. 
The decay (\ref{144}) 
has a small energy release ($E_{0}\simeq 18.6 \,~\rm{keV}$) and a
convenient time of life ($T_{1/2}\simeq 12.3$ years).

The standard effective Hamiltonian of the $\beta$-decay is given by
\be
\mathcal{H}_{I}^{\mathrm{CC}}= 
\frac{G_{F}}{\sqrt{2}}\,2 \bar e_{L}\gamma _{\alpha}\nu_{eL}\,~
j^{\alpha} + \mathrm{h.c.}\,.
\label{145}
\ee
Here $j^{\alpha}$ is the hadronic charged current,
$G_{F}$ is the Fermi constant and 
\be
\nu_{eL} = \sum_{i} U_{ei} \nu_{iL}\,,
\label{146}
\ee
where $\nu_{i}$ is the field of neutrino with mass  $m_{i}$
and $U$ is the unitary mixing matrix.

Neglecting the recoil of the final nucleus, from  { Eqs.}~(\ref{145}) 
and (\ref{146}) for the spectrum of electrons 
we obtain the following expression 
\be
\frac{d\,\Gamma}{d\,E}
= \sum_{i}|U_{ei}|^{2}\,~
\frac{d\,\Gamma_{i}}{d\,E}\,,
\label{147}
\ee
{ where}
\bea
&&\frac{d\,\Gamma_{i}}{d\,E} =
\label{148}\\
&&\,\, = C\,p\,(E+m_{e})\,(E_{0}-E)\,
\sqrt{(E_{0}-E)^{2}-m_{i}^{2}}\,F(E)\,\theta(E_{0}-E-m_{i})\,,
\nonumber
\eea
$E$ { being} the kinetic energy of the electron, $E_{0}$ the energy 
released in the decay and $ m_{e}$ is the mass of the electron. 
The Fermi function $F(E)$ takes into account the Coulomb interaction 
of the final particles and the constant $C$ is given by  the expression
$$C= G_{F}^{2}\frac{m_{e}^{5}}{2\,\pi^{3}}\,\cos^{2}\,\theta\, _{C}\,|M|^{2}
\,,$$ 
where $\theta\, _{C}$ is the Cabibbo angle { and}
$M$  the nuclear matrix element ({ which is} a constant).

The sensitivity to the neutrino mass of the present-day tritium experiments 
Troitsk~\cite{Troitsk} and Mainz~\cite{Mainz} is 2-3~eV. 
In the future experiment KATRIN~\cite{Katrin} { a} sensitivity { of} 
0.25~eV is expected. Taking into account that 
$\sqrt{\Delta m^{2}_{\rm{atm}}}\simeq 5\,10^{-2}\,\rm{eV}$ 
is much smaller than the sensitivities of the present and future 
tritium experiments, from { the relations} (\ref{143}) we can conclude 
that { the} neutrino  mass can be measured in these experiments only 
{ if the  neutrino mass spectrum is} practically degenerate:
$m_{1}\simeq m_{2}\simeq m_{3}$. { In this case,} from the unitarity 
of the mixing matrix we have:
\be
\frac{d\,\Gamma}{d\,E} = C\,p\,(E+m_{e})\,(E_{0}-E)\,
\sqrt{(E_{0}-E)^{2}-m_{1}^{2}}\,~F(E)\,.
\label{149}
\ee

Let us discuss now the results of the tritium experiments.
In the Mainz experiment~\cite{Mainz} the target is  molecular tritium
condensed on { a} graphite substrate. The spectrum of the electron 
is measured by the integral electrostatic spectrometer, which 
combines high luminosity with high resolution. 
The resolution of the { Mainz} spectrometer is equal to $4.8\,~\rm{eV}$. 
In the analysis of the experimental data four free variable parameters 
are used: the normalization $C$, the background $B$,
the released energy $E_{0}$ and the neutrino mass-squared $m^{2}_{1}$. 
The analysis of 1998, 1999 and 2001 data { gave the following result:}
\be
m^{2}_{1}= (-1.2 \pm 2.2 \pm 2.1)\,~\rm{eV}^{2}\,.
\label{150}
\ee
This value corresponds to the upper bound
\be
m_{1}< 2.2\,~\rm{eV}\,~~~ (95\%\,~ \rm{CL})\,.
\label{151}
\ee

The integral electrostatic spectrometer is  also used in the 
Troitsk neutrino experiment~\cite{Troitsk}: its resolution 
is 3.5-4~eV. In the Troitsk experiment the source is { a} gaseous 
molecular tritium source. From the four-parameter fit of the Troitsk data, 
for the parameter $m^{2}_{1}$ large negative values, in the range ($-10\div - 20$)~eV$^{2}$, were 
obtained. 
The investigation of the character of the measured spectrum suggests
that { the} effect of { obtaining a} negative $m^{2}_{1}$ is due 
to a step function superimposed on the integral continuous spectrum:
the step function in the integral spectrum corresponds to a narrow peak in
the differential spectrum.

In the analysis of the data, the authors of the Troitsk experiment 
added to the theoretical integral spectrum a step function with 
two additional variable parameters ({ the} position of the step 
$E_{\rm{step}}$  and { its} height). From the six-parameter fit of 
the data, the following value for the parameter $m^{2}_{1}$ 
\be
m^{2}_{1}= (-2.3 \pm 2.5 \pm 2.0)\,~\rm{eV}^{2}\,.
\label{152}
\ee
was found. From (\ref{152}) { the upper bound on}
 the neutrino mass $m_{1}$ 
\be
m_{1}< 2.2\,~\rm{eV}\,~~~( 95\%\,~\rm { CL})\,.
\label{153}
\ee
can be deduced.

The analysis of the Troitsk data shows that the position of the step 
$E_{0}-E_{\rm{step}}$ is periodically changed in the interval 5-15~eV and 
the average value of the height of the step is about $6\cdot 10^{-11} $. 
The existence of this anomaly was not confirmed by the Mainz 
experiment~\cite{Mainz}.

A new tritium experiment, KATRIN~\cite{Katrin}, is now under preparation.
In this experiment gaseous molecular source and frozen tritium source 
are planned to be used. The integral electrostatic spectrometer will have 
two parts: the pre-spectrometer, which will select electrons in the
last $\simeq 100$~eV of the spectrum, and the main spectrometer. 
{ The latter}  will have { a} resolution { of} $\simeq  1$~eV. 
It is expected  that the KATRIN experiment will start to collect data in
2007. After three years of running the accuracy of the measurement of the 
neutrino mass will reach $\simeq 0.25$~eV.

In order to understand the origin of the small neutrino masses we need to know
the nature of the massive neutrinos: are they Majorana or Dirac particles?
As we have seen, neutrino oscillation experiments can not answer
this fundamental question. The nature of the massive neutrinos can be revealed
in experiments on the search for neutrinoless double $\beta$-decay
of some even-even nuclei. { The next section will be devoted to
the discussion of} such experiments.

\section{Neutrinoless double $\beta$-decay}

The search for neutrinoless double $\beta$-decay
\be
(A,Z) \to (A,Z+2)+ e^{-}+ e^{-}
\label{154}
\ee
of some even-even nuclei is the most sensitive and direct way of 
{ investigating} the nature of neutrinos with definite masses.
The total lepton number in the process (\ref{154}) is violated and 
{ this} is allowed only if { the} massive neutrinos $\nu_{i}$ 
are Majorana particles.

We will assume that the Hamiltonian of the process has the standard 
form, Eq.(\ref{145}), and { that} the flavour field $\nu_{eL}$ 
is given by 
\be
\nu_{eL} = \sum_{i} U_{ei} \nu_{iL}\,,
\label{155}
\ee
where $\nu_{i}$ are Majorana fields.

The neutrinoless double $\beta$-decay ($(\beta\,\beta)_{0\nu}$ -decay)  
is { a process of} second order in the Fermi constant $G_{F}$,
with virtual neutrinos. For small neutrino masses { the} 
neutrino propagator is given by the expression
\be
<0|T(\nu_{eL}(x_{1})\nu^{T}_{eL}(x_{2}))|0> \simeq
m_{\beta\beta}\,~\frac{i}{(2\,\pi)^{4}}\int 
\frac{d^{4}p}{p^{2}}\, {e^{-ip(x_{1}-x_{2})}}\,
\frac{1-\gamma_{5}}{2}\,C\,,
\label{156}
\ee
where 
\be
m_{\beta\beta} = \sum_{i}U^{2}_{ei}\,m_{i}\,.
\label{157}
\ee

The total matrix element of the $(\beta\beta)_{0\nu}$ -decay is a product of
$m_{\beta\beta}$ and { the} nuclear matrix element, which does not 
depend on neutrino masses and mixing.

Results of many experiments on the search for $(\beta\beta)_{0\nu}$-decay
are available at present (see { Refs.}~\cite{Cremonesi,Zdesenko}).        
No indication in favour of $(\beta\beta)_{0\nu}$-decay { was} obtained 
so far.\footnote{ The recent claim~\cite{Klap01} of evidence of { a} 
$(\beta \beta)_{0\,\nu}$-decay, obtained from the reanalysis of the data 
of the Heidelberg-Moscow experiment, has been strongly criticized in  
{ Refs.}~\cite{FSViss02,bb0nu02}.}
The most stringent lower bounds for the time of life
of $(\beta\,\beta)_{0\,\nu}$-decay were obtained in the 
Heidelberg-Moscow~\cite{HM} and IGEX~\cite{IGEX}  $^{76}${Ge} experiments:
\bea
&&T^{0\nu}_{1/2}\geq 1.9 \cdot 10^{25}\, y\,~~ (90\% 
\,\rm{CL})\,~~ \rm{Heidelberg-Moscow}
\nonumber\\
&&T^{0\nu}_{1/2}\geq 1.57 \cdot 10^{25}\, y\,~~ (90\% 
\,\rm{CL})\,~~ \rm{IGEX}
\nonumber
\eea
Taking into account different calculations of the nuclear matrix element,
from these results the following upper bounds were obtained 
for the effective Majorana mass:
\be
|m_{\beta\beta}| \leq (0.35-1.24)\,~\rm{eV}\,. 
\label{158}
\ee
Many new experiments on the search for the neutrinoless 
double $\beta$-decay are in preparation at present (see 
{ Ref.}~\cite{Cremonesi}). In these experiments the sensitivities
$$|m_{\beta\beta}| \simeq (1\cdot10^{-1}-1.5\cdot 10^{-2})\,~\rm{eV}$$
are expected to be achieved.

The evidence for neutrinoless double $\beta$- decay would be a proof
that neutrinos with definite masses are Majorana particles.
The value of the effective Majorana mass $|m_{\beta\beta}|$  
combined with the values of the neutrino oscillation parameters,
obtained from the results of neutrino oscillation experiments,
would enable us to obtain important information about the character of
the neutrino mass spectrum, { the} minimal neutrino mass $m_1$ and 
{ the} Majorana CP phase (see { Ref.}~\cite{BGGKP} and 
references therein).

We will consider { here} three typical neutrino mass spectra.
\begin{enumerate}
\item
The hierarchy of neutrino masses $m_1 \ll m_2\ll m_3$.

In this case for the effective Majorana mass  we have the following 
upper bound :
\be
|m_{\beta\beta}| \leq \sin^{2} \theta_{\rm{sol}}\,\sqrt{\Delta m^{2}_{\rm{sol}}}+ 
|U_{e3}|^{2}\,\sqrt{\Delta m^{2}_{\rm{atm}}}.
\label{159}
\ee
Using the best-fit values of the oscillation parameters 
and the CHOOZ { limit} on $|U_{e3}|^{2}$ (see { Eqs.}~(\ref{046}), 
(\ref{069}) and (\ref{141})), we { obtain} from (\ref{159}) the { upper} bound
\be
|<m>| \leq 3.8 \cdot 10^{-3}\rm{eV}\,,
\label{160}
\ee
which is significantly smaller than the expected sensitivities 
of the future $(\beta \beta)_{0\nu}$- experiments.

\item
Inverted hierarchy of neutrino masses: $m_{1}\ll m_{2}<m_{3}$.

The effective Majorana mass is given, in this case, by the expression
\be
|m_{\beta\beta}|\simeq  
\left( 1 - \sin^{2}2\,\theta_{\rm{sol}}\,\sin^{2}\alpha \right)
^{\frac{1}{2}}\,\sqrt{\Delta m^{2}_{\rm{atm}}},
\label{161}
\ee
where $\alpha=\alpha_{3}-\alpha_{2}$ is the the difference of the Majorana 
CP phases ($U_{ei}= |U_{ei}|\,~e^{i\,\alpha_{i}}$).
From this expression it follows that
\be
\sqrt{\Delta m^{2}_{\rm{atm}}}
\,~ |\cos2\,\theta_{sol}|\lesssim |m_{\beta\beta}|
\lesssim \sqrt{\Delta m^{2}_{\rm{atm}}}\,,
\label{162}
\ee
where the upper and lower bounds correspond to the 
case of  CP conservation with equal and opposite CP parities 
of $\nu_{3}$ and $\nu_{2}$, { respectively}.

Using the best-fit value of the parameter $\tan^{2}\theta_{sol}$
(see { Eq.}~(\ref{119})), we have
\be
\frac{1}{2}\,\sqrt{\Delta m^{2}_{\rm{atm}}}
\lesssim |m_{\beta\beta}|
\lesssim\sqrt{\Delta m^{2}_{\rm{atm}}}\,,
\label{163}
\ee
Thus, in the case of the inverted mass hierarchy, the scale of 
$|m_{\beta\beta}|$ is determined by $\sqrt{\Delta m^{2}_{\rm{atm}}}$.
Should the value of $|m_{\beta\beta}|$ be in the range (\ref{163}), 
(which can be reached in the future experiments on the search for  
$(\beta \beta)_{0\nu}$-decay), then we would have an argument 
in favour of { the} inverted neutrino mass hierarchy.

\item
Practically degenerate neutrino mass spectrum:
$m_2 \simeq m_3\simeq m_1\gg \sqrt{\Delta m^{2}_{\rm{atm}}}$.

The effective Majorana mass in this case is given by the expression
\be
|m_{\beta\beta}| \simeq m_{1}\,~|\sum_{i=1}^{3}U_{ei}^{2}|.
\label{164}
\ee
Neglecting { the} small contribution of 
$|U_{e3}|^{2}$ (~$|U_{e1}|^{2}$ in the case of the inverted hierarchy),
for $|m_{\beta\beta}|$ one obtains relations 
Eqs.~(\ref{161})-(\ref{163}) in which $\sqrt{\Delta m^{2}_{\rm{atm}}}$ is
replaced by $ m_1$.  Thus, a signature of the degenerate neutrino 
mass spectrum { would be the occurrence of a value }
$|m_{\beta\beta}| \gg \sqrt{\Delta m^{2}_{\rm{atm}}}$.

It is obvious that for neutrino mass 
$m_{1}$ we have the following bound
\be
|m_{\beta\beta}| \leq m_{1}\leq\frac{|m_{\beta\beta}|}
{|\cos 2\,\theta_{\rm{sol}}|}\simeq 2\,|m_{\beta\beta}|
\label{165}
\ee
and the parameter $\sin^{2}\alpha$, which characterizes the 
violation of the CP invariance in the lepton sector~\cite{BGGKP}, 
is given by the relation
\be
\sin^{2}\alpha \simeq \left( 1 -\frac{|m_{\beta\beta}|^{2}}{m_{1}^{2}}
\right)\,~\frac{1}{\sin^{2}2\theta_{\rm{sol}}}\,.
\label{166}
\ee

{ Thus,} if the mass $m_1$ is measured in the future experiments 
and the value of the parameter $\sin^{2}2\theta_{\rm{sol}}$ 
is determined with high accuracy in the
KamLAND~\cite{Kamland}, BOREXINO~\cite{BOREXINO} and other neutrino 
experiments, from the results of the future $(\beta \beta)_{0\nu}$ 
experiments  information on the Majorana CP phase can be inferred.
\end{enumerate}

All previous conclusions are based on the assumption that the 
value of the effective Majorana mass $|m_{\beta\beta}|$ can be 
obtained from the measurement of the life-time of the 
$(\beta \beta)_{0\nu}$-decay. 
However, the determination of the parameter $|m_{\beta\beta}|$
from the experimental data requires the knowledge of the nuclear matrix elements.
At present there are large uncertainties in the calculation of these
quantities (see, for example, { Refs.}~\cite{Faessler,Suhonen,Elliott}).
Different calculations of the lifetime of the $(\beta \beta)_{0\,\nu}$-decay 
differ by about one order of magnitude.

In { Ref.}~\cite{BGr} a method was proposed, which allows one 
to check the results of the calculations  
of the  $(\beta \beta)_{0\nu}$-decay nuclear matrix elements 
in a model independent way. { This method can be applied }
if $(\beta \beta)_{0\nu}$-decay of {\it different} nuclei is observed.

\section{Conclusion}

{ About forty years after the original idea of B. Pontecorvo,}
compelling evidence in favour of neutrino oscillations were obtained 
 in the S-K~\cite{S-K}, SNO~\cite{SNO,SNOCC,SNONC}, 
KamLAND~\cite{Kamland} and other neutrino experiments.
These findings opened a new field of research in 
particle physics and astrophysics: {\em the physics of massive and mixed 
neutrinos}.

From the results of the experiments it follows that neutrino masses are
many orders of magnitude smaller than the masses of leptons and quarks. 
Tiny neutrino masses are a first signature of a new physics, 
beyond the Standard Model.

There { remain} many unsolved problems in the physics of massive and 
mixed neutrinos. The problem { concerning} $|U_{e3}|^{2}$ is an urgent one:
on the value of $|U_{e3}|^{2}$ depends the possibility to study the effects 
of the three-neutrino mixing and in particular { the}
effects of  CP violation in the lepton sector.

Another problem is the { one connected with the LSND 
experiment}~\cite{LSND}. If the LSND result will be confirmed by the 
MiniBOONE experiment~\cite{MiniB}, { this will imply} that the number 
of light neutrinos is more than three and in addition to the three 
flavour neutrinos sterile neutrino(s) must exist. 
If, { on the contrary,} LSND result is refuted, the minimal scheme 
with three massive and mixed neutrinos will be a plausible possibility.

The problem of the nature of massive neutrinos (Dirac or Majorana?)
is the most fundamental one. This problem can be solved by the
experiments on the search for neutrinoless double $\beta $-decay.
From the existing data the following bound on the effective Majorana 
mass was found: $|m_{\beta\beta}| \leq (0.3-1.3)$~{eV}.
In future  experiments, now in  preparation, { a} significant 
improvement of the sensitivity is expected. 

{ Due to} the interference nature of the neutrino oscillation phenomenon
and { to the} possibilities of exploring large values
of ${L}/{E}$, neutrino oscillation experiments are sensitive to very 
small values of { the} neutrino mass-squared differences.
The determination of the absolute value of { the} neutrino masses 
requires  the so-called direct measurements and it is  
a challenging problem. From { the} data of the tritium experiments 
the { upper} bound $m_{1}\leq 2.2\,\rm{MeV}$ was obtained. 
The future experiment KATRIN~\cite{Katrin} 
is expected to be sensitive to $m_{1}\simeq 0.25$~{MeV}.  

The progress in  cosmological measurements,
achieved in the last years, allows one to reach
$\simeq 1$~eV sensitivity { for} the sum of { the} neutrino 
masses $\sum_{i}m_{i}$. The most stringent bound, 
$$\sum_{i}m_{i} < 0.71~ \rm{eV}\,,$$
was recently obtained in { Ref.}~\cite{WMAP}.
For { a} degenerate neutrino mass spectrum this bound implies
$$m_{i} < 0.23~ \rm{eV}\,.$$
In Ref.~\cite{WMAP} the 2dF Galaxy Redshift Survey~\cite{2dF} data, 
recent WMAP high precision data and other cosmological data were used.
{ A} significant improvement of this { limit}
is expected with the future Sloan Digital Sky Survey~\cite{SDS}, future
WMAP and other data.

We have discussed here mainly { the} 
phenomenology of neutrino mixing and oscillations, 
and the 
most recent experimental data. For models of neutrino masses and 
mixing { we refer the reader to the recent} 
reviews~\cite{AlFer,King,Mohapatra}.

\end{document}